\UseRawInputEncoding
\documentclass[
 aip,
 amsmath,amssymb,
 reprint,
]{revtex4-2}
\usepackage{graphicx}
\usepackage{dcolumn}
\usepackage{bm}
\usepackage{mathptmx}
\usepackage{xcolor}
\usepackage{hyperref}
\usepackage{cancel}
\usepackage{etoolbox}
\usepackage{siunitx}
\usepackage{subcaption}
\usepackage{ragged2e}
\DeclareCaptionJustification{justified}{\justifying}
\newcommand\Eq[1]{Eq.~(\ref{#1})}
\newcommand\rr{\mathbf{r}}

\newcommand\EE{\mathbf{E}^\omega(\rr)}
\newcommand\HH{\mathbf{H}^\omega(\rr)}
\newcommand\DD{\mathbf{D}^\omega(\rr)}
\newcommand\BB{\mathbf{B}^\omega(\rr)}

\newcommand\dd{\mathbf{p}^\omega(\rr)}
\newcommand\mm{\mathbf{m}^\omega(\rr)}
\newcommand\inddipoles{\begin{bmatrix}\dd\\\mm\end{bmatrix}}
\newcommand\fields{\begin{bmatrix}\EE\\\HH\end{bmatrix}}
	\newcommand\aee{\alpha_{\mathrm{ee}}^\omega}
\newcommand\aem{\alpha_{\mathrm{em}}^\omega}
\newcommand\ame{\alpha_{\mathrm{me}}^\omega}
\newcommand\amm{\alpha_{\mathrm{mm}}^\omega}
\newcommand\etaee{\eta_{\mathrm{ee}}}
\newcommand\etaem{\eta_{\mathrm{em}}}
\newcommand\etame{\eta_{\mathrm{me}}}
\newcommand\etamm{\eta_{\mathrm{mm}}}
\newcommand\alphamat{\underline{\underline{\alpha}}^\omega}
\newcommand\den{1-\baromega^2-\ii\bargamma\baromega}
\newcommand\varepsilonr{\varepsilon_{\mathrm{r}}}
\newcommand\mur{\mu_{\mathrm{r}}}
\newcommand\varepsiloneff{\varepsilon_{\mathrm{eff}}^\omega}
\newcommand\mueff{\mu_{\mathrm{eff}}^\omega}
\newcommand\kappaeff{\kappa_{\mathrm{eff}}^\omega}

\newcommand\ii{\mathrm{i}}
\newcommand\baromega{\bar{\omega}}
\newcommand\bargamma{\bar{\gamma}}

\newcommand\hatV{\hat{V}}

\newcommand\rhoL{\rho_{\mathrm{L}}}
\newcommand\rhoD{\rho_{\mathrm{D}}}

\newcommand{\upsub}[1]{\sb{\mathrm{#1}}}
\newcommand{\upsup}[1]{\sp{\mathrm{#1}}}

\begingroup\lccode`~=`_\lowercase{\endgroup\let~\upsub}
\begingroup\lccode`~=`^\lowercase{\endgroup\let~\upsup}

\AtBeginDocument{
  \catcode`_=12 \catcode`^=12
  \mathcode`_="8000
  \mathcode`^="8000
}

\emergencystretch=\maxdimen
\begin{document}

\title{On enhanced sensing of chiral molecules in optical cavities}

\author{Philip Scott}
\email{philip.scott@kit.edu}
\affiliation{Institute of Applied Physics, Karlsruhe Institute of Technology, 76128 Karlsruhe, Germany}
\author{Xavier Garcia-Santiago}
\affiliation{Institute of Theoretical Solid State Physics, Karlsruhe Institute of Technology, 76128 Karlsruhe, Germany}
\affiliation{JCMWave GmbH, 14050 Berlin, Germany}
\author{Dominik Beutel}
\affiliation{Institute of Theoretical Solid State Physics, Karlsruhe Institute of Technology, 76128 Karlsruhe, Germany}
\author{Carsten Rockstuhl}
\affiliation{Institute of Theoretical Solid State Physics, Karlsruhe Institute of Technology, 76128 Karlsruhe, Germany}
\affiliation{Institute of Nanotechnology, Karlsruhe Institute of Technology, 76021 Karlsruhe, Germany}
\author{Martin Wegener}
\affiliation{Institute of Applied Physics, Karlsruhe Institute of Technology, 76128 Karlsruhe, Germany}
\affiliation{Institute of Nanotechnology, Karlsruhe Institute of Technology, 76021 Karlsruhe, Germany}
\author{Ivan Fernandez-Corbaton}
\email{ivan.fernandez-corbaton@kit.edu}
\affiliation{Institute of Nanotechnology, Karlsruhe Institute of Technology, 76021 Karlsruhe, Germany}

\date{\today}

\begin{abstract}
The differential response of chiral molecules to incident left- and right- handed circularly polarized light is used for sensing the handedness of molecules. Currently, significant effort is directed towards enhancing weak differential signals from the molecules, with the goal of extending the capabilities of chiral spectrometers to lower molecular concentrations or small analyte volumes. Previously, optical cavities for enhancing vibrational circular dichroism have been introduced. Their enhancements are mediated by helicity-preserving cavity modes which maintain the handedness of light due to their degenerate TE and TM components. In this article, we simplify the design of the cavity, and numerically compare it with the previous one using an improved model for the response of chiral molecules. We use parameters of molecular resonances to show that the cavities are capable of bringing the vibrational circular dichroism signal over the detection threshold of typical spectrometers for concentrations that are one to three orders of magnitude smaller than those needed without the cavities, for a fixed analyte volume. Frequency resolutions of current spectrometers result in enhancements of more than one order (two orders) of magnitude for the new (previous) design. With improved frequency resolution, the new design achieves enhancements of three orders of magnitude. We show that the TE/TM degeneracy in perfectly helicity preserving modes is lifted by factors that are inherent to the cavities. More surprisingly, this degeneracy is also lifted by the molecules themselves due to their lack of electromagnetic duality symmetry, that is, due to the partial change of helicity during the light-molecule interactions.
\end{abstract}
\maketitle
\section{Introduction and summary}
The fundamental biomolecules in living organisms are chiral. Therefore, the effect that the two enantiomers of a chiral molecule have on them can be very different\cite{Nguyen2006}. Sensing the dominant handedness of chiral molecules is thus an important task in chemistry, biology, and pharmacology. Two of the phenomena that are exploited for this task are the differential absorption of circularly polarized light, known as circular dichroism (CD), and the polarization rotation of linearly polarized light, known as optical rotation (OR). The CD, like the OR, is equal in magnitude but opposite in sign for similar ensembles of opposite enantiomers\cite{Barron2004}. The chirality of the molecules manifests itself when electromagnetic radiation induces electronic molecular transitions, molecular vibrations, or molecular rotations. The three different mechanisms are activated by illumination frequencies in the near-UV, infrared, and the \SI{}{\giga\hertz} region, respectively\cite{Nafie2011}. In all three cases, the inherent weak chiral response of the molecules limits the usability of existing spectrometers for measuring the CD signal as the molecular concentration and/or the analyte volume decrease. Enhancing the differential signal is then necessary. The photonics community has taken on this challenge with a recent stream of theoretical and experimental studies \cite{Muller2000,Muller2002,Auguie2011,Schaeferling2012,Bougas2012,Hentschel2012,Valev2013,Patterson2013,GarciaEtxarri2013,Wu2014,Bougas2014,Sofikitis2014,Yoo2015,Nesterov2016,Schaeferling2016,Cameron2016,GarciaEtxarri2017,Zhao2017,Vazquez-Guardado2018,Sofikitis2018,Hanifeh2018b,Mohammadi2018,Poulikakos2018,Garcia-Guirado2018,Forbes2019,Graf2019,Feis2019,Neufeld2019,Ayuso2019,Solomon2019,Garcia-Guirado2020,Droulias2020,Semnani2020,Lasa2020,Takumi2020,Solomon2020}. 

 \begin{figure*}[ht!]
    
    \hspace*{-0.5cm}
    \begin{subfigure}[hb!]{.45\linewidth}
    \centering
    \includegraphics[width=7.5cm, height=7.5cm]{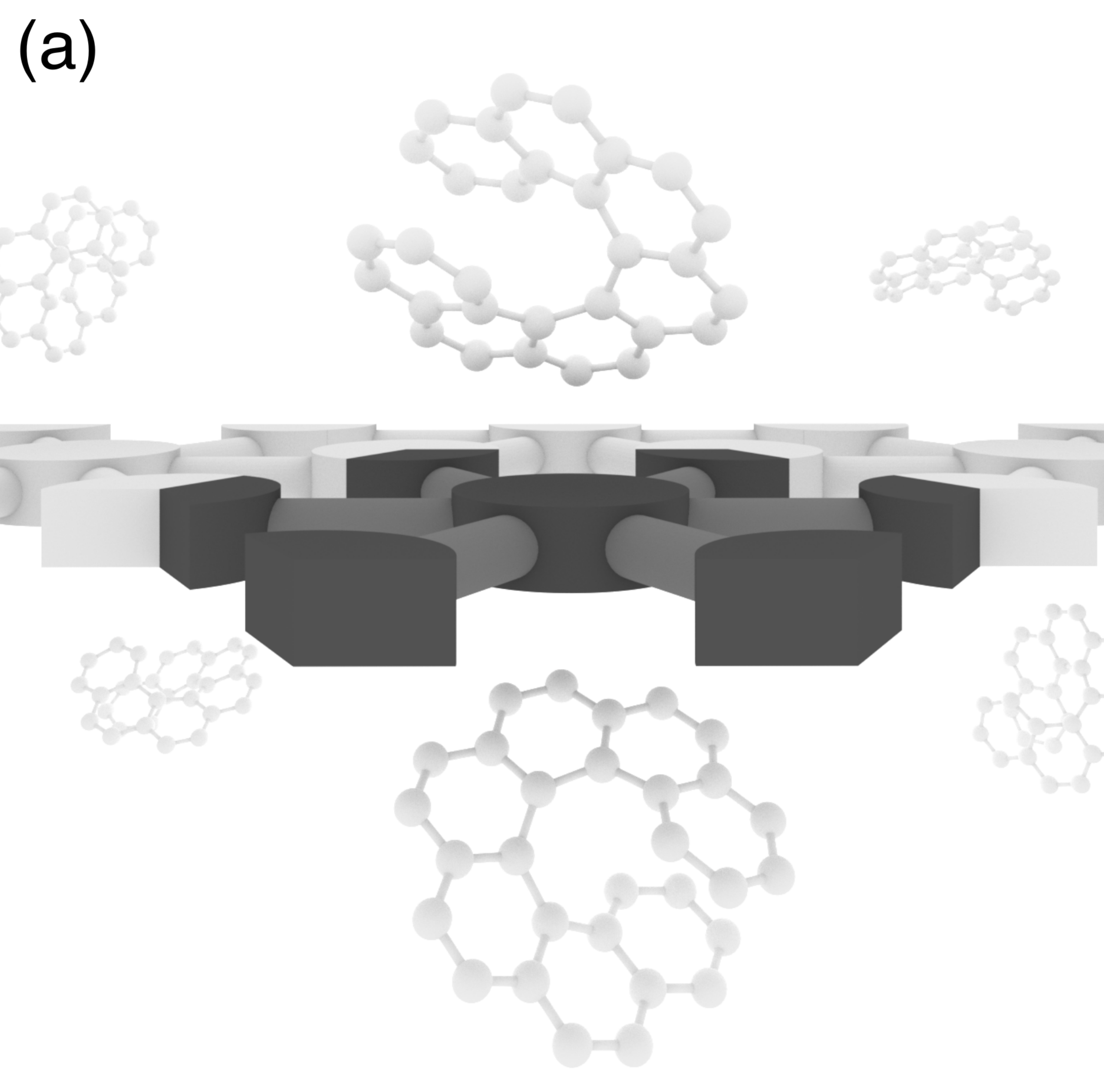}
	\end{subfigure}
	\begin{subfigure}[hb!]{.45\linewidth}
    \centering
    \includegraphics[width=7.5cm, height=7.5cm]{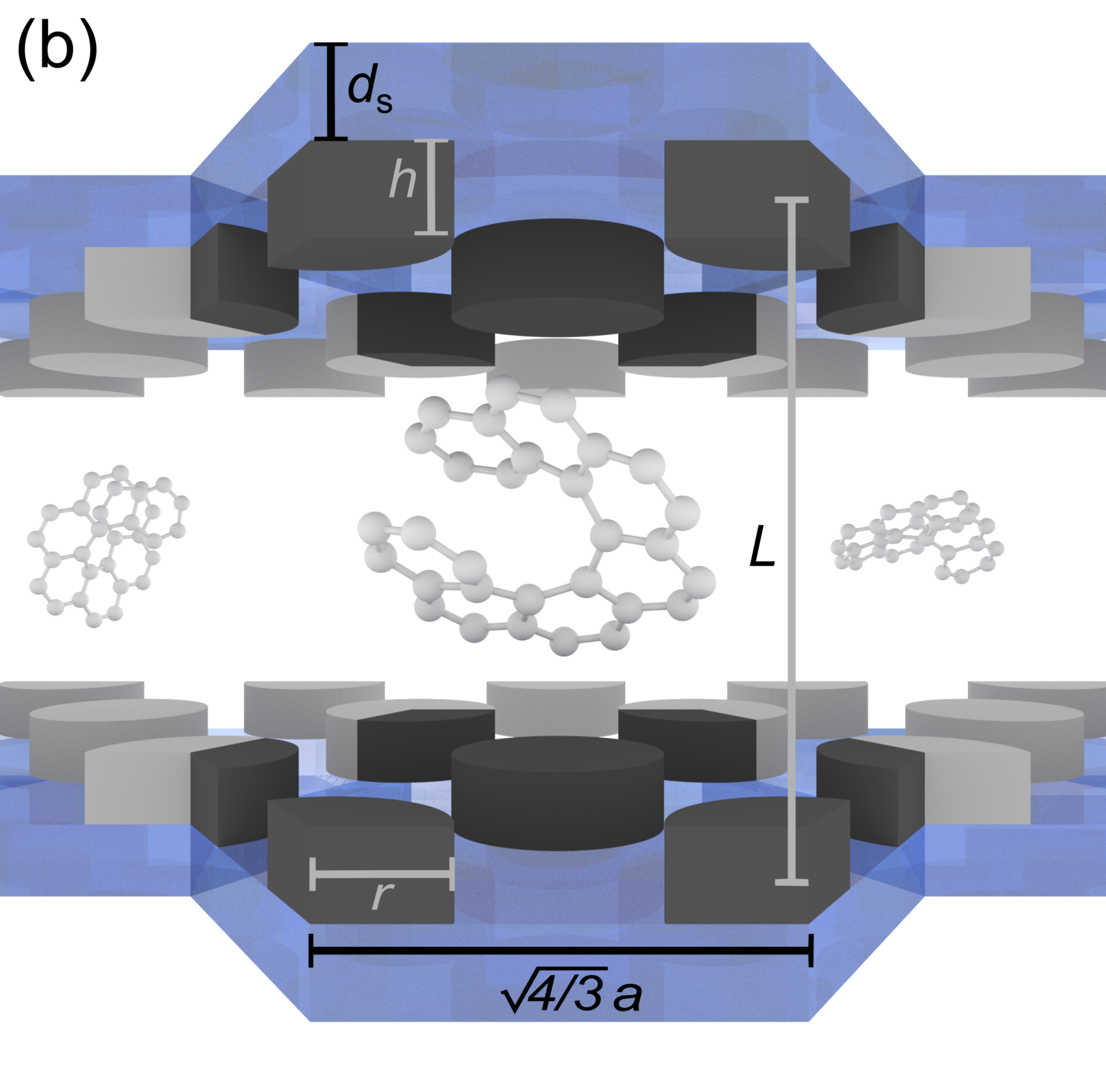}
    \end{subfigure}

    \hspace*{-0.5cm}
    \begin{subfigure}[hb!]{.45\linewidth}
    \centering
    \includegraphics[width=7.5cm, height=7.5cm]{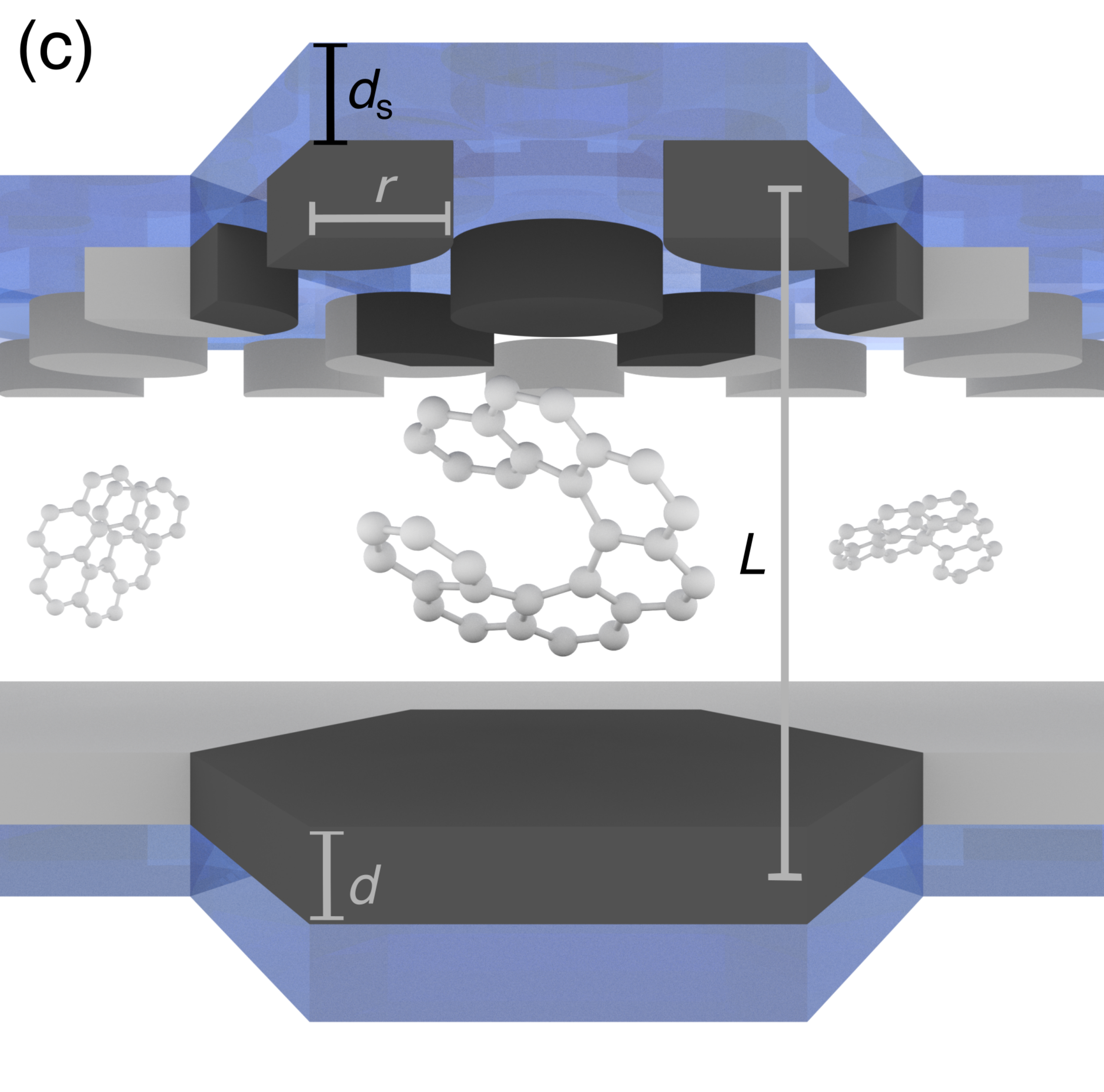}
	\end{subfigure}
	\begin{subfigure}[hb!]{.45\linewidth}
    \centering
    \includegraphics[width=7.5cm, height=7.5cm]{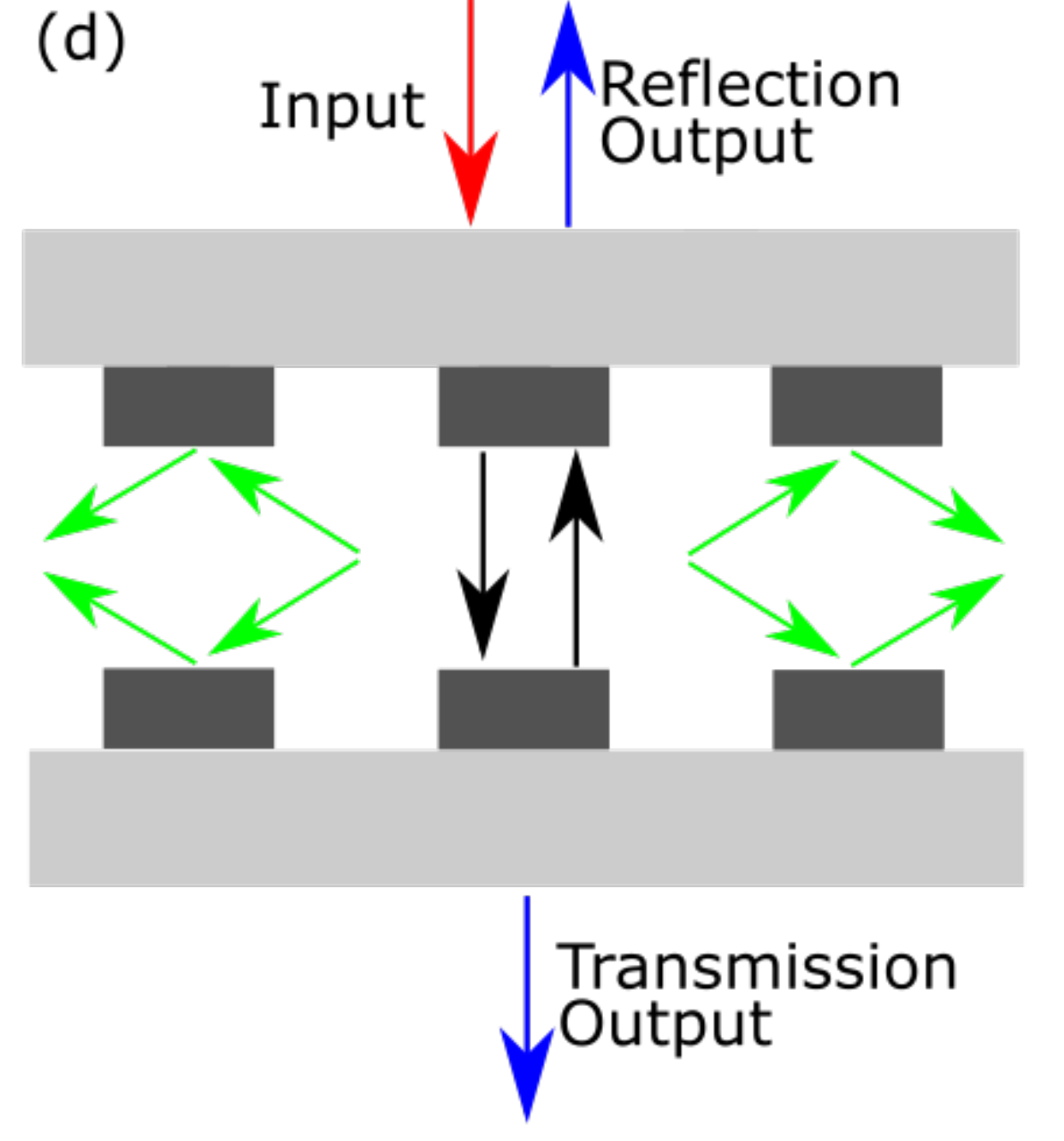}
	\end{subfigure}
    
	 \caption{\label{fig:single} (a-c) Sketches of systems for CD enhancement. (a) A single-array of silicon disks connected by rods, without a substrate. (b) A cavity formed by two silicon disk arrays arranged in hexagonal lattices where $r=$\SI{1.92}{\micro\meter}, $h=$\SI{1.32}{\micro\meter}, and $a=$\SI{5.76}{\micro\meter}, with substrate thickness $d_{s}=$\SI{2}{\micro\meter}, and cavity length $L$. (c) A cavity formed by an array of silicon disks with the same parameters as in (b), and a homogeneous slab of silicon with thickness $d=$\SI{1}{\micro\meter}. (d) Ray diagram showing the path of the light as it passes through the double-array cavity depicted in (b). The behavior on the single-array cavity depicted in (c) is essentially the same. The light impinges from above in this depiction (red). The zeroth diffraction orders of the two arrays excite Fabry-Perot modes in the cavity (black), which are not useful for CD enhancement. The CD enhancing modes are due to the first diffraction orders, which produce modes of large transverse momentum (green). In these modes, the light inside the cavity bounces without changing handedness at grazing angles between the two disk arrays until it is diffracted out of the cavity by one of the arrays (blue). The two arrays are assumed to be periodic in the lateral directions. }
\end{figure*}

\begin{figure*}[ht!]
    \includegraphics[width=\linewidth]{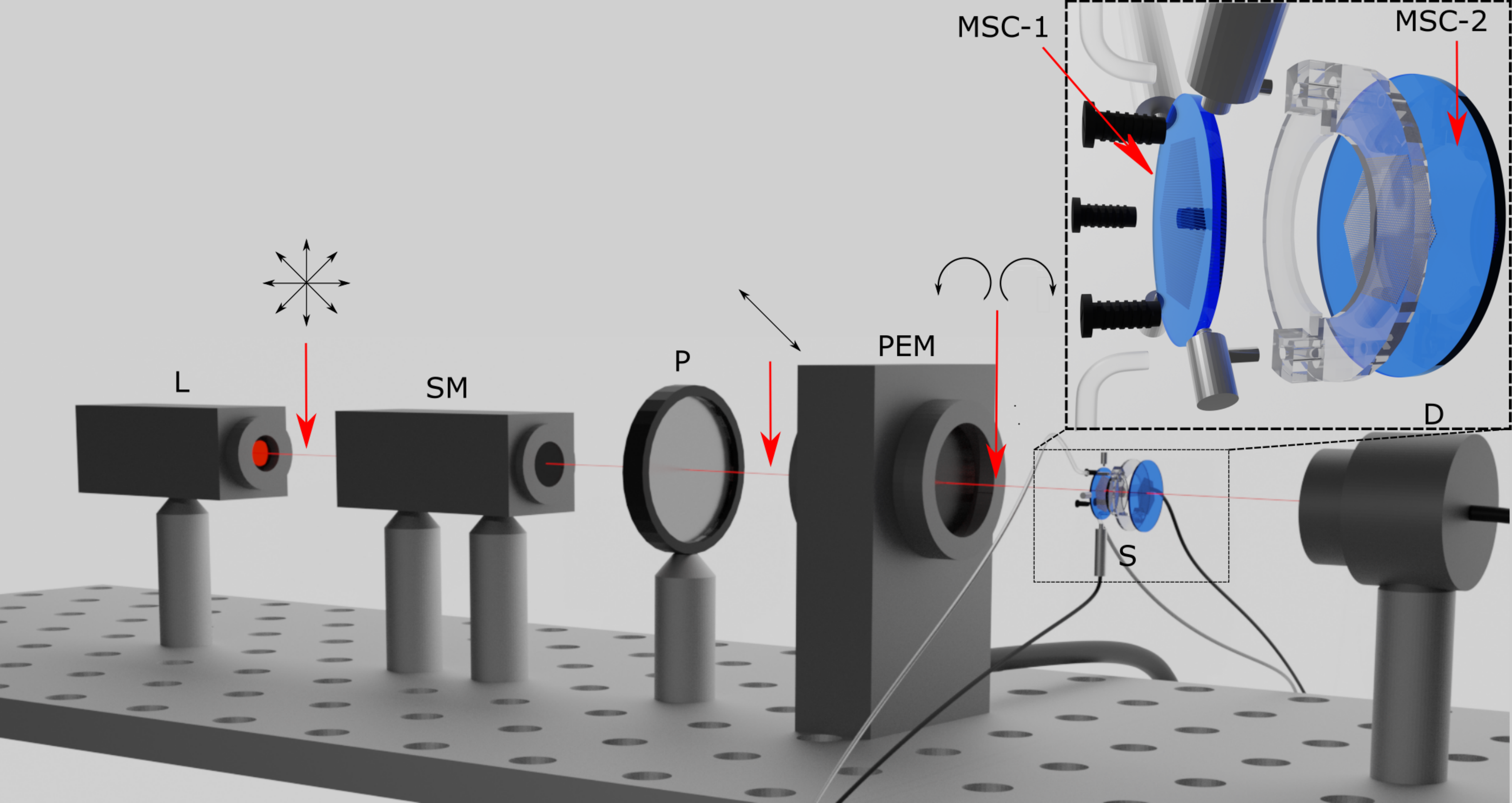}
	\caption{\label{fig:CDspec} An artistic scheme of the components of a CD spectrometer setup with our CD enhancing cavity surrounding the sample. An enlarged version of this later part of the setup can be seen in the inset, where the blue disks indicate the substrates with the metasurface mirrors on top of them. MSC-1: Metasurface mirror 1, MSC-2: Metasurface mirror 2, L: Light source, SM: Spectrometer, P: Linear polarizer, PEM: Photoelastic modulator, S: Sample encased in our CD enhancing cavity, D: Detector. The light emerges from the light source, enters a spectrometer, and then travels through a linear polarizer. Next, the photoelastic modulator continuously flips the light between left- and right-handed circularly polarized-light at a certain modulation frequency. The light interacts with the sample and finally hits the detector, which sends the information to a computer for analysis. The red arrows mark several points in the light path for which the polarization of light is depicted with black arrows.}
\end{figure*}

Enhanced sensitivity for the measurement of optical rotation has been experimentally shown in Cavity Ring Down Polarimetry (CRDP) setups\cite{Muller2000,Muller2002,Bougas2012,Bougas2014,Sofikitis2014}, where light pulses pass through the sample multiple times, and quarter-wave plates and/or Faraday rotators are used to overcome previously encountered difficulties\cite{Poirson1998}. In such setups, the single-pass optical pathlength is on the order of one meter. At much smaller scales, one of the main strategies is to increase the light-molecule interaction by means of photonic micro-structures that resonantly enhance the fields. In this context, the electromagnetic helicity\cite{Zwanziger1968,Calkin1965,Deser1976,Birula1981,Birula1996,Afanasiev1996,Trueba1996,Drummond1999,Coles2012,FerCor2012p,Cameron2012,Bliokh2013,Cameron2013,Nieto2015,Gutsche2016,FerCor2016,Elbistan2017,Andrews2018,Vazquez2018,Hanifeh2018b,Crimin2019,FernandezGuasti2019,Poulikakos2019,Bernabeu2019,FerCor2019b} has proven to be a useful quantity in the analysis and design of systems for enhanced sensing of chiral molecules\cite{Graf2019,Feis2019}. The electromagnetic helicity can be seen as the generalization of the concept of circular polarization to general electromagnetic fields, including near fields, evanescent fields, and cavity modes. Achiral and resonant systems that do not change the helicity of the incident light can be seen as the resonant version of typical chiro-optical spectrometers\cite{Graf2019}. Achirality is needed to avoid signal distortions including non-zero signals from achiral analytes \cite{Wu2014,Nesterov2016,Mohammadi2018,Graf2019}. While helicity preserving resonances are not a must, they are optimal under some conditions\cite{Graf2019} for enhancing the CD signal. Several achiral systems featuring helicity preserving resonances for enhanced sensing of chiral molecules have been reported\cite{Solomon2019,Graf2019,Feis2019,Lasa2020,Vazquez-Guardado2018,Droulias2020,Solomon2020}. Helicity preservation is achieved if and only if the TE and TM responses of the system are identical (see Chap.~2.4 in Ref.~\onlinecite{FerCorTHESIS}). For resonant systems, this implies the degeneracy of TE and TM modes. Very recently, we have reported the theoretical design of an optical cavity featuring helicity preserving modes for enhancing the vibrational circular dichroism (VCD) signal of chiral molecules\cite{Feis2019}. The cavity, sketched in Fig.~\ref{fig:single}(b), consists of two parallel mirrors made from arrays of high permittivity dielectric disks. The disks are placed on top of substrates and are arranged in a periodic hexagonal pattern. The two mirrors are placed parallel to each other at distances on the order of tens of micrometers, thereby forming a cavity that is filled by the solution of chiral molecules. In contrast to many other designs where the CD enhancements occur only in close proximity to some structured photonic material, the differential absorption in our proposed cavity is enhanced practically across the whole volume of the analyte inside the cavity. The modal fields of almost pure helicity corresponding to the first diffraction orders of the cavity provide average enhancements of the VCD signal of over two orders of magnitude for cavity lengths of a few tens of micrometers. This, together with the intrinsic spectral tunability of a cavity make the design in Ref.~\onlinecite{Feis2019} very appealing for experimental realization. Figure~\ref{fig:CDspec} depicts a possible experimental setup. At this point, the refinement of the theoretical models becomes necessary for obtaining precise performance predictions, and the consideration of alternative designs of increased operational simplicity is warranted.

In this article, we compare two optical cavity designs for CD enhancement: Our previous design featuring two identical arrays of tailored silicon disks, which we will call ``double-array cavity'', and a new simplified design where one of the arrays is substituted by a thin, homogeneous silicon slab, which we will call ``single-array cavity''. This seemingly minor modification constitutes a very valuable simplification experimentally because it eliminates the need for lateral alignment of the two silicon-disk arrays on a deep sub-micrometer scale. The performance of the cavities is accurately predicted by using a realistic model for the optical response of the solution of chiral molecules. The parameters of the model can be fixed using existing spectroscopic measurements of the targeted resonance for the desired chiral molecule. While we will here focus on molecular solutions in the liquid phase, the same modeling strategy can be used for the gas phase. In particular, the model accounts for the duality symmetry-breaking of the molecules\cite{FerCor2012p,FerCor2013}, that is, for the partial change of electromagnetic helicity during light-molecule interaction. We show that the degree of duality breaking increases as the Kuhn's dissymmetry factor \cite{Kuhn1930} ($g$) for the resonance decreases. This symmetry breaking, which is completely inconsequential in typical CD spectrometers, can affect the modes of the cavity due to the long light-matter interaction times. We show that the ideal TE/TM degeneracy in perfectly helicity preserving modes is disturbed by factors that are inherent to the design of the cavities, and also by the lack of electromagnetic duality symmetry of the chiral molecules. These disturbances cause a splitting of the TE and TM modes. As the splitting grows, the enhancement decreases and the spectral signatures of the CD signals progressively change from a single helicity preserving resonance to two sharper helicity preserving features separated by a helicity flipping region, in which the sign of the signal changes. The contribution of the molecules to the splitting increases as the molecular concentration divided by the Kuhn's dissymmetry factor of the molecular resonance increases, and as the linewidth of the spectral features of the cavity response decreases. With these effects taken into account, we show that both cavity designs can provide VCD enhancements of more than two orders of magnitude, which we exemplify for a particular resonance of the binol molecule. For the double-array cavity, the standard frequency resolution of current spectrometers is much finer than the width of the enhancement lines. For the single-array cavity, the linewidth of the helicity-preserving features that achieve enhancements of three orders of magnitude are comparable to the standard frequency resolution of commercial VCD spectrometers. The single-array cavity can also be operated exploiting much wider features that achieve more than one order of magnitude of negative enhancement, where the sign of the CD signal is deterministically changed. For the simplest operation of the cavities when targeting a particular resonance, the concentration should be low enough to avoid the more severe effects of the splitting and large enough so that the CD signal is above the sensitivity threshold of the measurement apparatus. The upper limit in the concentration could be avoided by the use of parameter estimation techniques exploiting that the effects of the TE/TM splittings in the CD signal can be accurately reproduced using a simple analytical model. Our results imply that the effects of the duality breaking of the molecules on the response of helicity-preserving optical cavities are less pronounced for electronic CD and more pronounced for rotational CD, with vibrational CD lying in the middle. This statement follows from the connection between duality symmetry breaking and Kuhn's dissymmetry factor, and the known scaling of $g$ for each kind of CD: The largest values of $g$ in electronic, vibrational, and rotational CD spectroscopy are of the order $10^{-1},10^{-3},\text{ and } 10^{-5}$, respectively\cite{Mason1963,Jasco2011,Salzman1991}, and the typical values are at least an order of magnitude lower in each case. The rest of the article is organized as follows. 

In Sec.~\ref{sec:one}, we make the idealized assumption that the molecules do not change the helicity of the light that they interact with and study the undisturbed response of the two cavities. We identify the different causes of TE/TM splitting that degrade the helicity preserving properties of the cavity modes and change the spectral shapes of the enhancement lines. We show that these changes can be modeled by the composition of a TE and a TM resonance with varying frequency detuning. In Sec.~\ref{sec:two} we develop a model for the constitutive relations of the solution of a chiral molecule for the frequency region containing a molecular resonance. In Sec.~\ref{sec:three} we use the model to target a particular resonance of binol in its VCD spectrum, and to study the effects of the duality breaking of the molecules on the two cavities. Section~\ref{sec:four} contains the concluding remarks.

\section{Cavity Design Comparison\label{sec:one}}

\begin{figure*}[ht!]
    \hspace*{-0.7cm}
    \begin{subfigure}[hb!]{.45\linewidth}
    \centering
    \includegraphics[width=8.2cm, height=6cm]{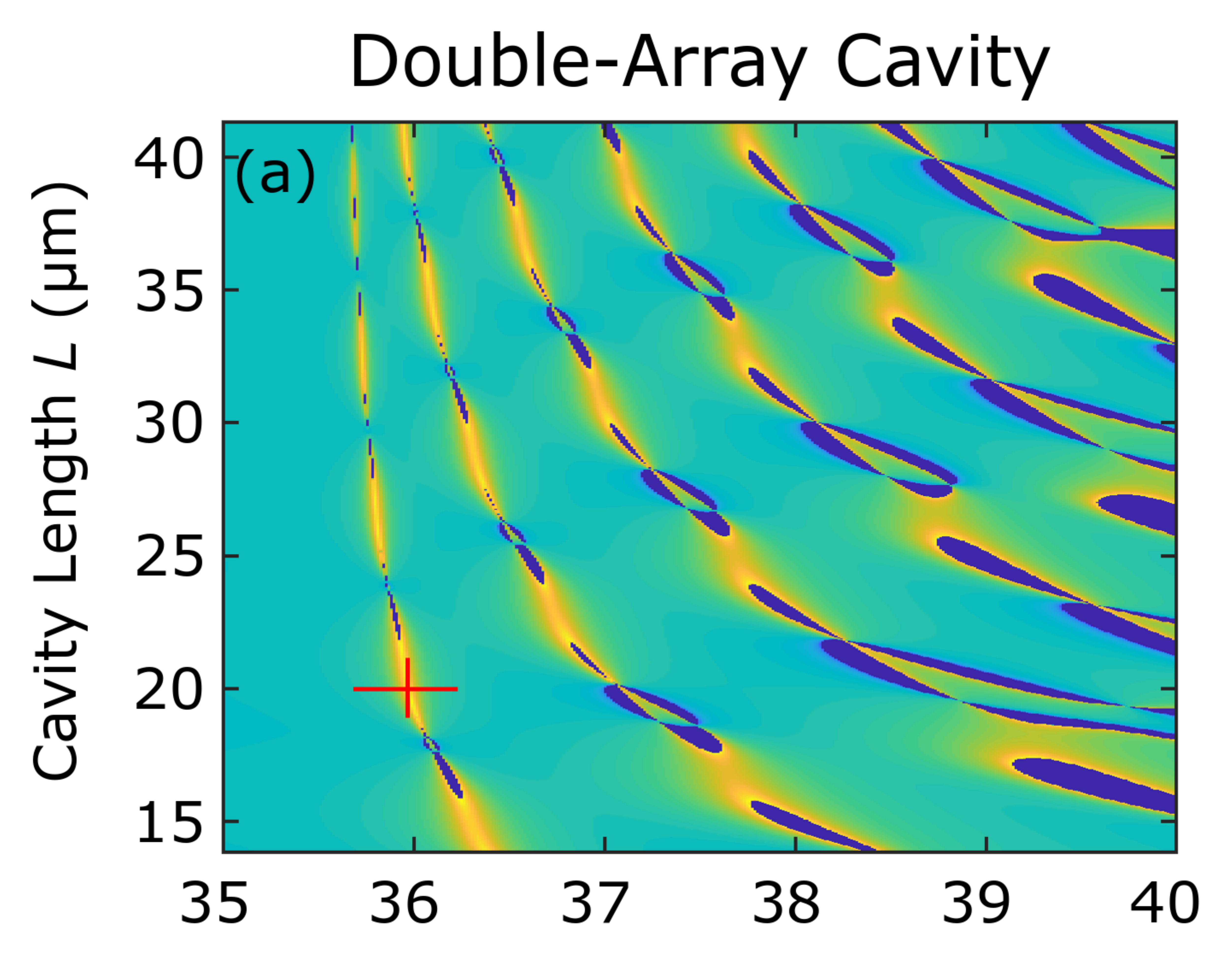}
	\end{subfigure}
	\begin{subfigure}[hb!]{.45\linewidth}
    \centering
    \includegraphics[width=9.05cm, height=6cm]{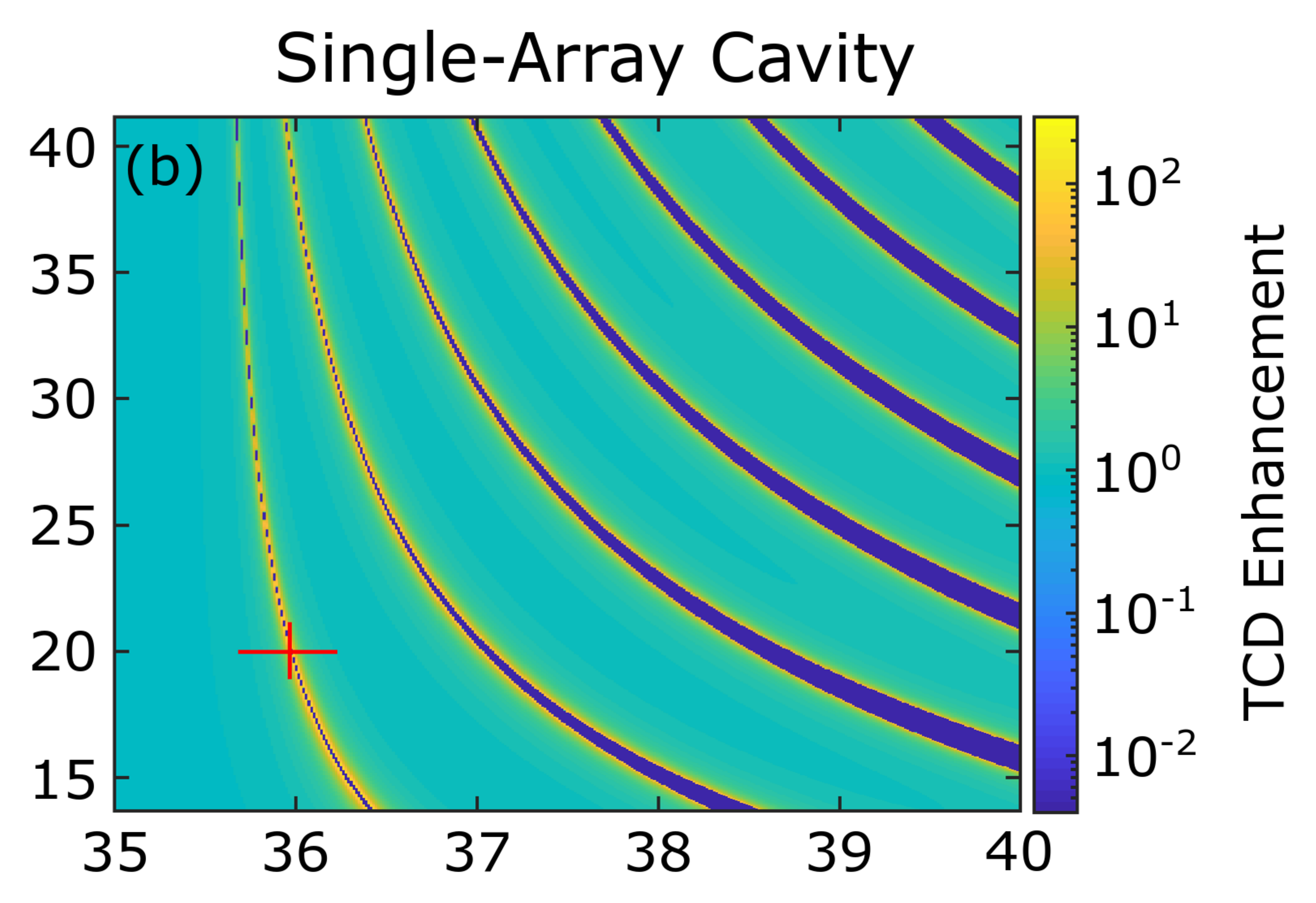}
    \end{subfigure}

    \hspace*{-0.5cm}
    \begin{subfigure}[hb!]{.45\linewidth}
    \centering
    \includegraphics[width=7.8cm, height=5.8cm]{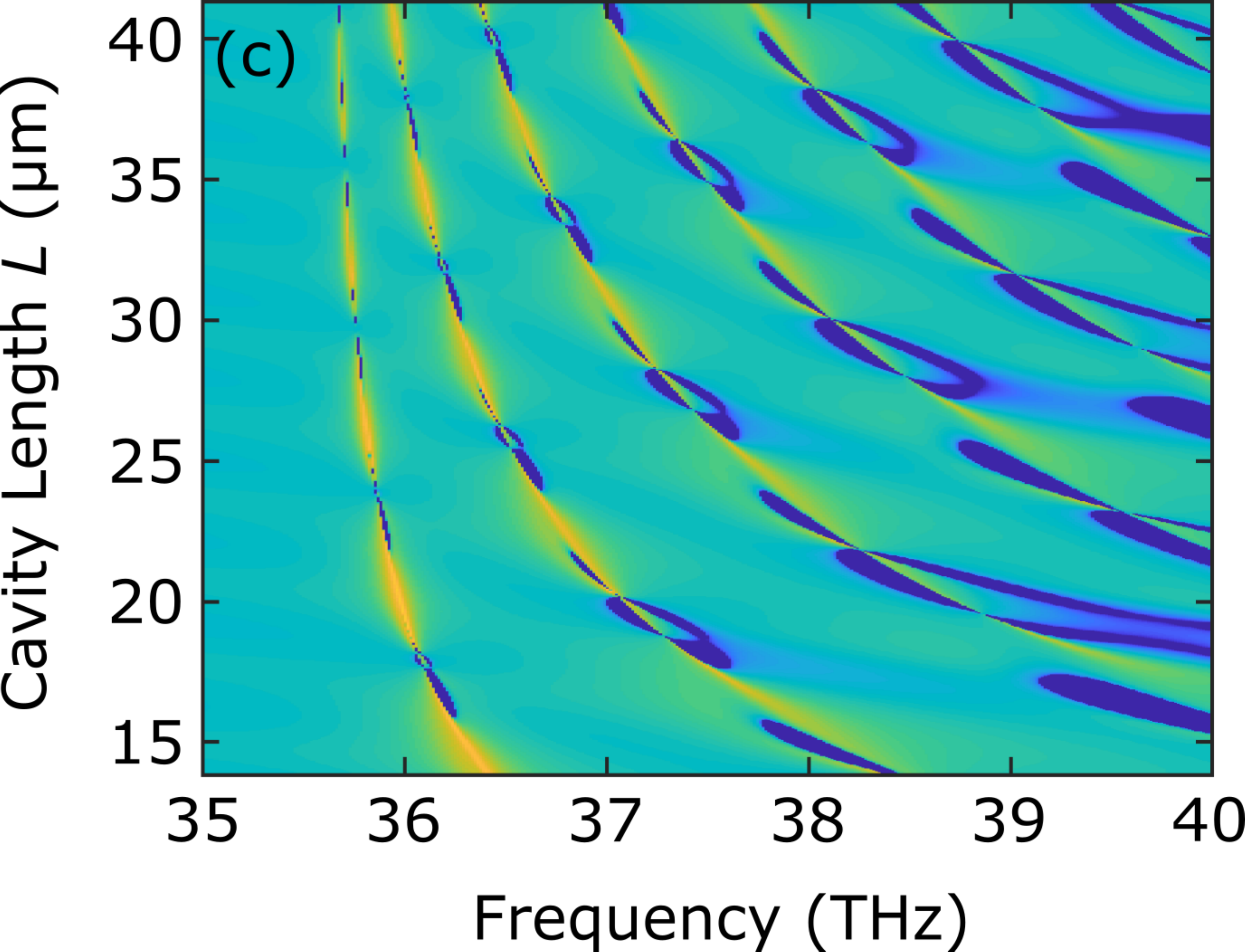}
	\end{subfigure}
	\begin{subfigure}[hb!]{.45\linewidth}
    \centering
    \includegraphics[width=8.8cm, height=5.8cm]{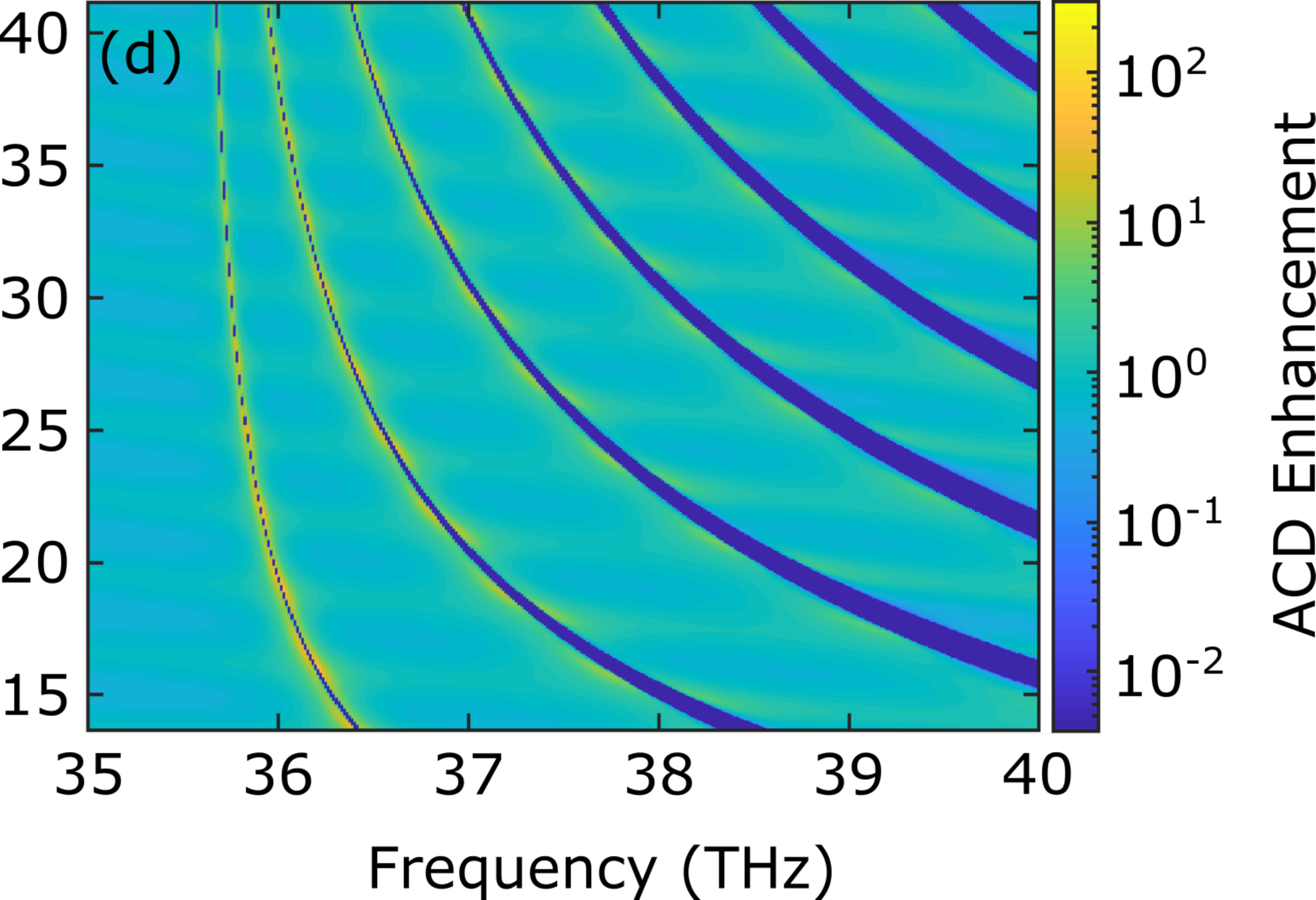}
	\end{subfigure}
	\caption{\label{fig:double} Comparison plots of the double-array cavity and the single-array cavity. Panels (a) and (c) show the TCD and ACD enhancements of the double-array cavity, respectively. Panels (b) and (d) show the TCD and ACD enhancements of the silicon slab and array cavity, respectively. The false color scale has been truncated from below at the level of 10${^{-2.5}}$ in linear units, and there exist negative enhancement values in some of the dark blue regions. The red crosses mark the data cuts shown in Fig.~\ref{fig:compar}. }
\end{figure*}

We consider the three systems depicted in Fig.~\ref{fig:single}. A single planar hexagonal array of silicon disks connected by rods (without a substrate), a cavity formed by two parallel hexagonal arrays of silicon disks, and a cavity formed by an array of silicon disks and a homogeneous silicon slab. The not-to-scale molecule drawings in Fig.~\ref{fig:single} represent the solution of chiral molecules. The disk arrays and the silicon slab defining the cavities are placed on substrates, and the medium outside the two cavities is air. The systems are sequentially illuminated from the top by perpendicularly incident plane waves of opposite polarization handedness. 

The geometrical parameters of the single planar array in Fig.~\ref{fig:single}(a) can be optimized to produce a resonant and helicity preserving response upon illumination at normal incidence\cite{Solomon2019,Graf2019}. The lattice pitch $a\sqrt{4/3}$ is set to ensure that only the zeroth diffraction order is allowed to propagate through the solution. At the resonance frequency, the incident circularly polarized light produces strong near fields of pure helicity attached to the disks. These near-fields result in CD enhancement factors that reach $\approx 20$ at the surface of the disks, but then drop quickly with increasing distance from the disks. Therefore, only the molecules in the immediate vicinity of the disks experience a significant effect. While evaporation of chiral molecules onto the silicon disks\cite{Garcia-Guirado2020} provides a means to exploit the near-field effect, the practical applicability of single-array sensors will be limited by the fact that the CD signal would not be enhanced over macroscopic volumes of analyte. This problem is solved by the double-array cavity in Fig.~\ref{fig:single}(b), where the helicity preserving resonances are achieved in a quite different way \cite{Feis2019}. As illustrated in Fig.~\ref{fig:single}(d), the arrays of silicon disks and their lattice pitch are designed such that circularly polarized light impinging onto the cavity under normal incidence is diffracted into a first order towards large angles approaching 90 degrees while preserving the helicity of the light. Next, when the cavity resonance condition is met for the diffracted light, the light inside the cavity bounces obliquely between the two disk arrays at grazing angles, leading to a large interaction time with the chiral molecules. Crucially, the helicity is preserved because the TE- and TM-polarizations of light become nearly degenerate in their behavior for grazing incidence onto the disk arrays. Finally, the light is diffracted out of the cavity in a normal direction by the silicon disk arrays and can be detected. Importantly, the strong modal fields of almost pure helicity inside the cavity are not attached to the disks, but are spread across the entire volume between the two silicon disk arrays. Resonant enhancement factors exceeding a factor of 2000 at some points inside the cavity can be reached. When averaging the enhancement factors across the volume of the entire cavity, enhancement factors exceeding 100 at cavity lengths of $\approx\SI{20}{\micro \metre}$ have been numerically demonstrated in Ref.~\onlinecite{Feis2019} for idealized molecules with a large degree of helicity preservation. For the experimental realization of the cavity, the need for precise lateral alignment of the two silicon disk arrays must be emphasized: Misalignments result in an overall chiral system which distorts and can overwhelm the desired signal from the molecules (see Ref.~\onlinecite[App.~IV]{Feis2019}). In practice, such alignment can pose a substantial challenge. This difficulty motivates an alternative new design for the cavity, depicted in Fig.~\ref{fig:single} (c), where one of the two arrays is replaced by an thin, homogeneous silicon film that acts as a regular mirror under conditions of grazing incidence. The idea behind the simplification is that only one silicon disk array is needed for coupling of the internal cavity modes featuring grazing angles to the external perpendicular directions of illumination and measurement. Since the new design has only a single disk array, it does not need lateral alignment. Both designs still need the two sides of the cavity to be as parallel as possible to each other. In our simulations, we have assumed that the two sides are perfectly parallel. In practice, this can be controlled by using a set of three linear actuators in a triangular formation at the edges of the cavity to adjust one of the sides (see the inset in Fig.~\ref{fig:CDspec}). In our simulations, we also assume that the arrays and the silicon slab extend to infinity in the lateral directions. In practice, the lateral extent should be large enough so that the edge effects are negligible.

\begin{figure*}[ht!]
	\begin{subfigure}[t]{.47\linewidth}
		\includegraphics[width=1.05\textwidth]{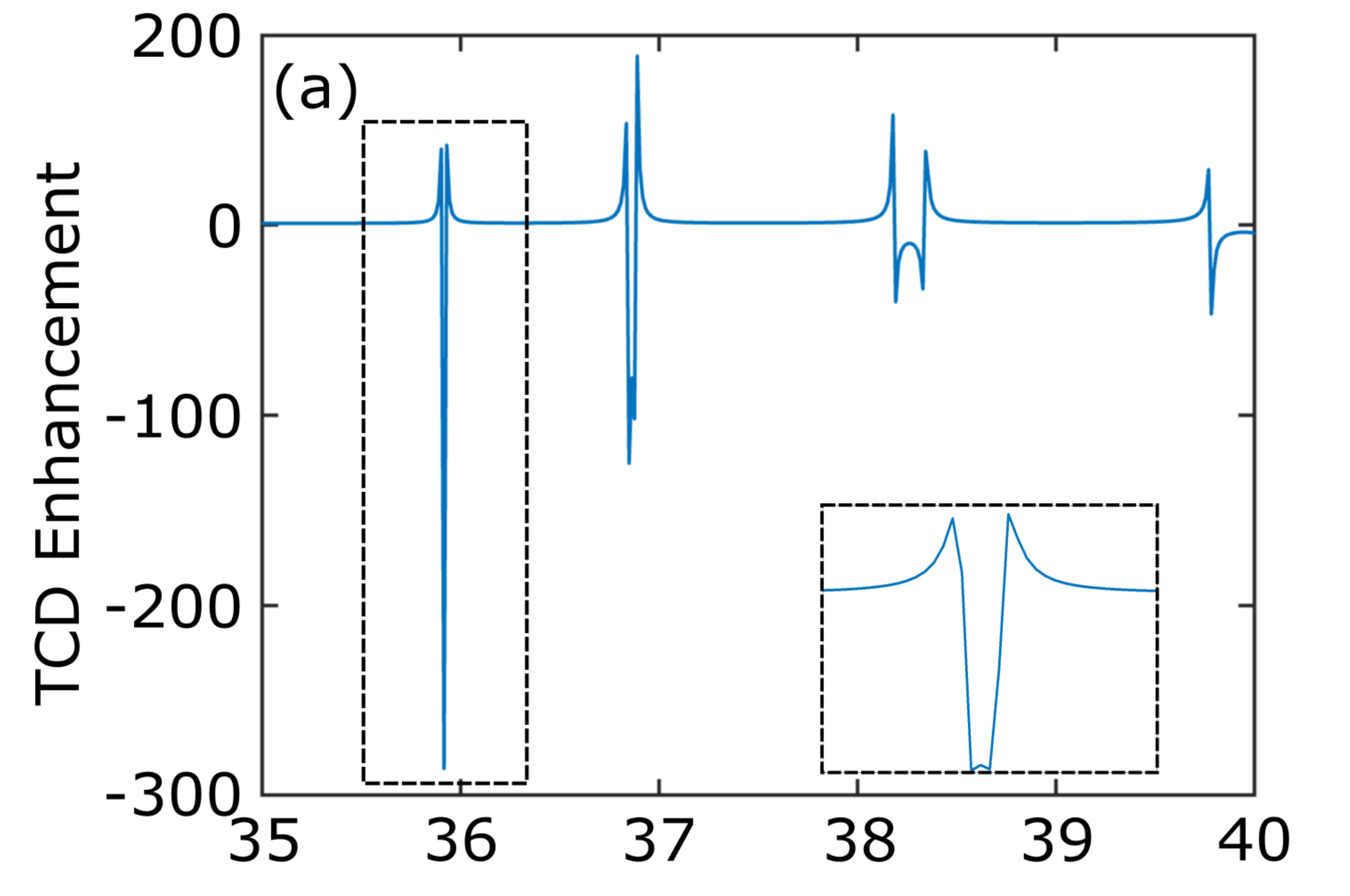}
	\end{subfigure}
	\hspace{0.1cm}
	\begin{subfigure}[t]{.47\linewidth}
		\includegraphics[width=0.98\textwidth]{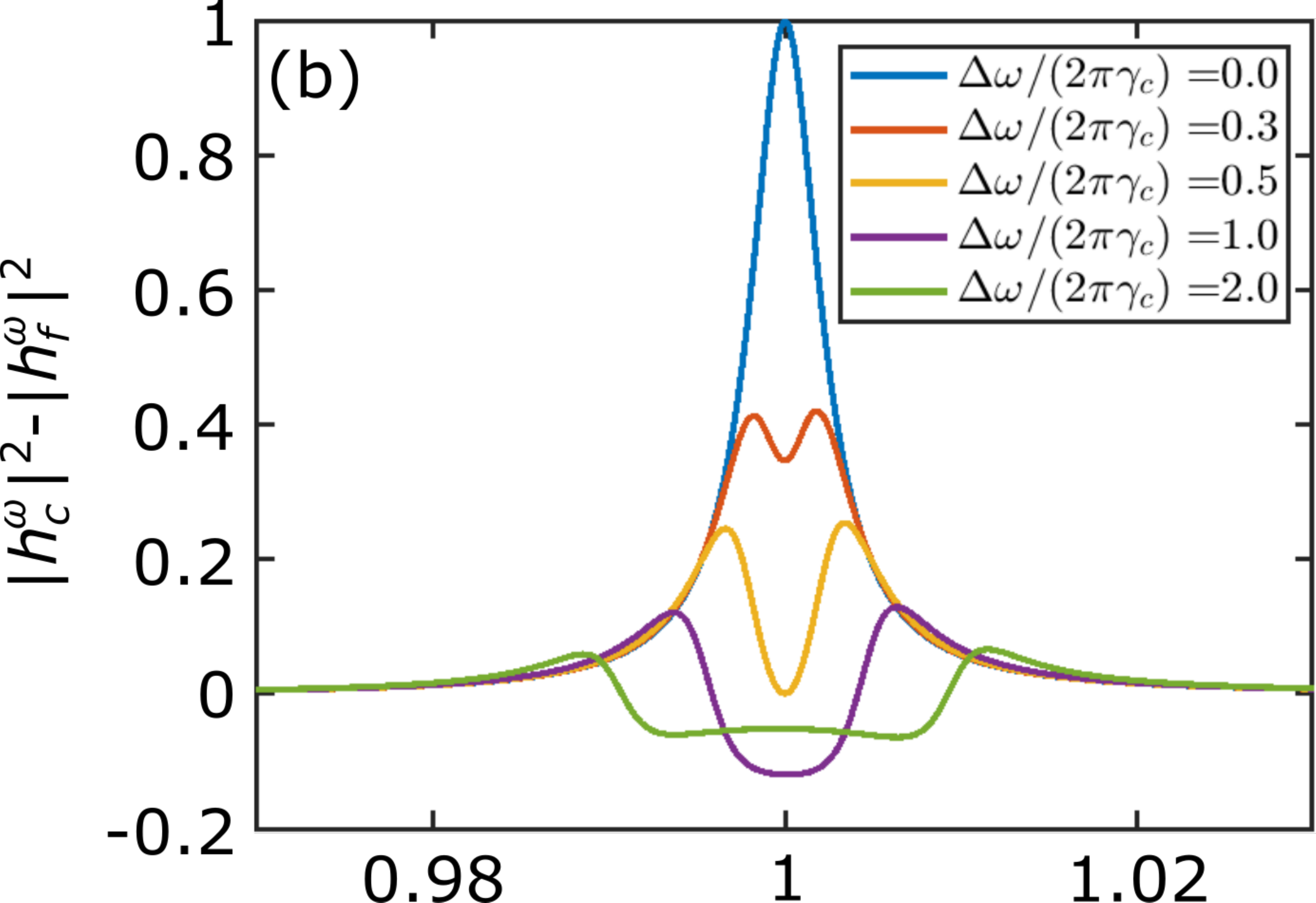}
	\end{subfigure}
	\\
	\begin{subfigure}[t]{.47\linewidth}
		\includegraphics[width=1.05\textwidth]{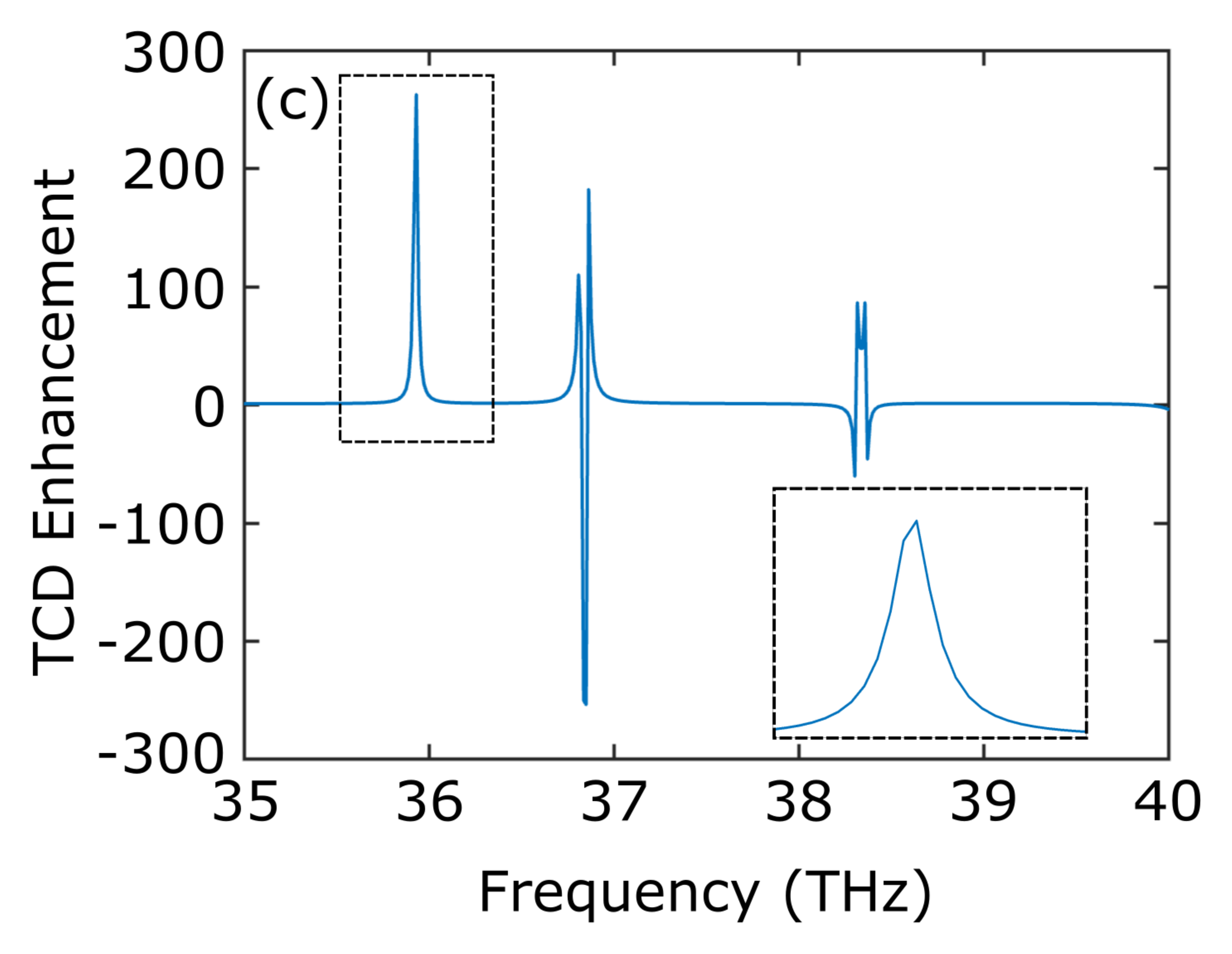}
	\end{subfigure}
	\hspace{0.1cm}
	\begin{subfigure}[t]{.47\linewidth}
		\includegraphics[width=1.02\textwidth]{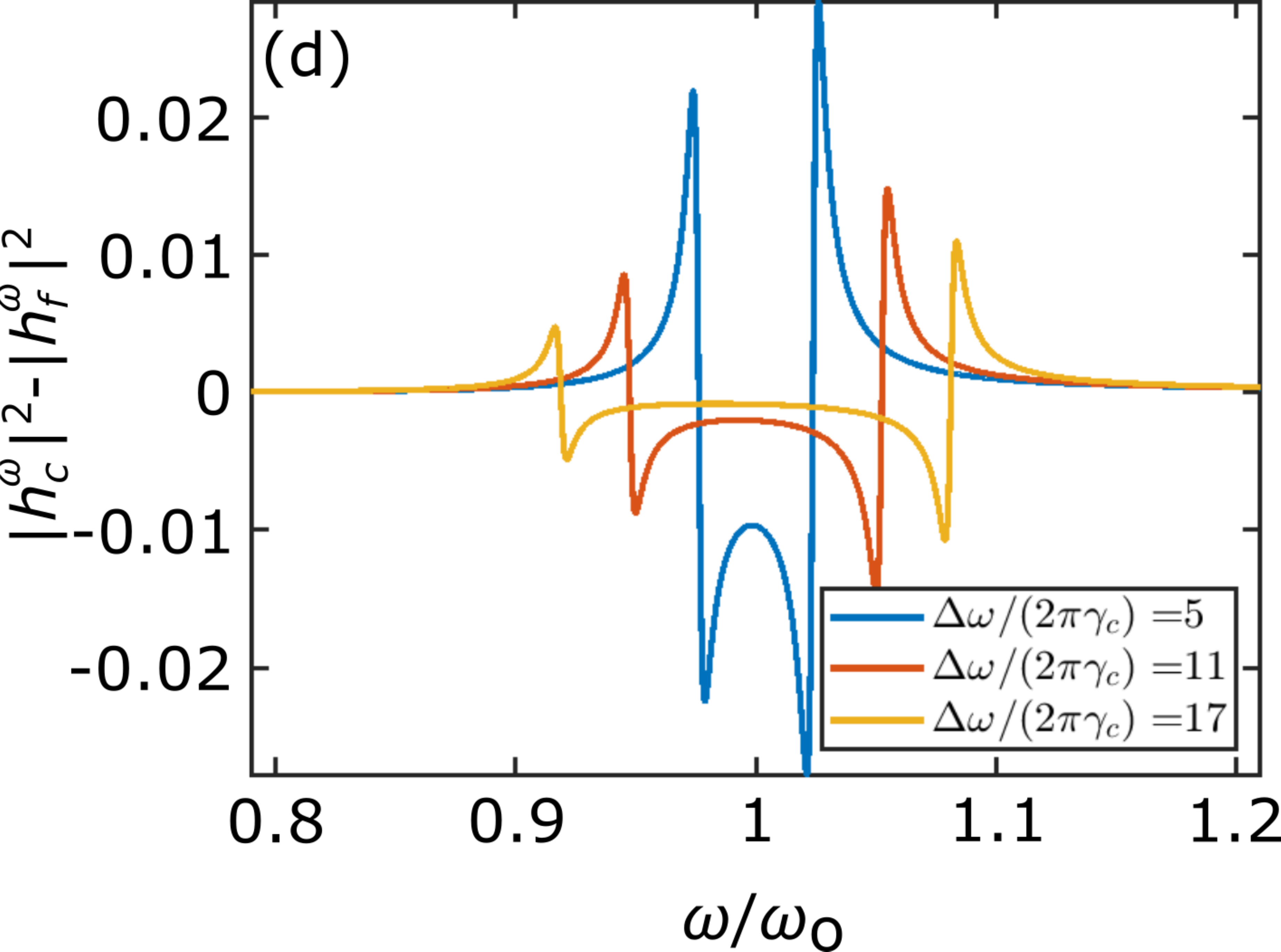}
	\end{subfigure}
	\caption{\label{fig:gotit}Panels (a) and (c) show the TCD enhancement of the single- and double-array cavity, respectively, as a function of frequency for a cavity length $L$=\SI{21.52}{\micro\meter}. The TCD enhancement is shown here in linear units. In each case, the insets show the lowest frequency mode in more detail. Panels (b) and (d) show the normalized helicity content resulting from a model where a TE and a TM resonance with linewidths equal to $\gamma_c$ are located $\Delta f$ \SI{}{\hertz} apart (see text).}
\end{figure*}
We now explore the simplified single-array cavity design and compare it to the double-array cavity. We use the same T-matrix-based simulation of laterally infinitely periodic arrays\cite{Beutel2020} that we used in Ref.~\onlinecite{Feis2019} for numerical evaluations. The T-matrices of the silicon disks are obtained with the help of JCMsuite\cite{Garcia2018} assuming that the relative permittivity of silicon is 11.9. We also assume that the substrates and the solvent have a relative permittivity equal to 2.14. The solution of chiral molecules is modeled using the Condon-Tellegen constitutive relations for monochromatic fields with an implicit $\exp(-\ii\omega t)$ time dependence
\begin{equation}
	\label{eq:cons}
	\begin{split}
		\DD&=\varepsiloneff\EE+\ii\frac{\kappaeff}{c_0}\HH,\\
		\BB&=\mueff\HH-\ii\frac{\kappaeff}{c_0}\EE,\
	\end{split}
\end{equation}
where SI units are used and we assume for now a non-dispersive response of the molecules: $\varepsiloneff=2.14(1+\ii10^{-4})\varepsilon_0$, $\mueff=(1+\ii10^{-4})\mu_0$, and $\kappaeff=\ii 10^{-4}$, where $\epsilon_0$ ($\mu_0$) is the vacuum permittivity (permeability). We note that the assumed values meet the equality $Z_{\text{sol}}^2\varepsiloneff=\mueff$, where $Z_{\text{sol}}$ is the optical impedance of the solvent. This implies that the molecules have duality symmetry\cite{FerCor2013}, that is, that they do not change the helicity of the light that they interact with. We will later analyze the effect of removing this idealization. Figure~\ref{fig:double} shows the absorption CD (ACD) and transmission CD (TCD) enhancements for both cavities as a function of the frequency $f$ in units of \SI{}{\tera\hertz} and cavity length $L$ in \SI{}{\micro\meter}. The definition of ACD is $\mathrm{ACD} = \frac{P_+ - P_-}{2P_0}$, where $P_+$ and $P_-$ are the outgoing powers for the subsequent left-handed (``+'') and right-handed (``-'') circularly polarized incident light, assuming identical incident powers equal to $P_0$. The outgoing power is the sum of the powers of the transmitted and reflected fields. We recall that the array of disks is diffracting in the solvent but not in air, so only the forward and backward directions perpendicular to the plane of the arrays can send energy outside of the cavity ($P_{\pm}=P_\pm^{\text{fwd}}+P_\pm^{\text{bwd}}$). The definition of TCD is 
\begin{equation}
	\text{TCD}=\frac{P^{\text{fwd}}_{+}-P^{\text{fwd}}_{-}}{P^{\text{fwd}}_{+}+P^{\text{fwd}}_{-}},
\end{equation}
where only the transmitted powers $P^{\text{fwd}}_{\pm}$ need to be detected. The ACD (TCD) enhancement factor is defined by the ACD (TCD) signal with the enhancing cavity divided by that without the enhancing cavity. We note that the false color scale of Fig.~\ref{fig:double} has been truncated from below at the level of 10${^{-2.5}}$ in linear units, and that there exist negative enhancement values in some of the dark blue regions, which we analyze later. 

Figure~\ref{fig:double} shows very similar positions of the high-enhancement lines for the two cavities. As explained in Ref.~\onlinecite{Feis2019}, the lines in the $(L,f)$ parameter space correspond to different cavity modes excited by the first diffraction orders of the hexagonal lattice upon normal incidence. The position of the enhancement lines depends on several parameters (see Eq.~(B8) in Ref.~\onlinecite{Feis2019}): The cavity length $L$, the mode number, the lattice spacing $a$, and the refractive indices of the disks and the solvent. These parameters can be modified to tune the frequency response of the cavities and enhance the CD of a given chiral molecule resonance. While the position of the lines is essentially the same in both cavities, their structure is different. One difference is that the enhancement along a given line is more uniform for the single-array cavity. The enhancement variations along the double-array cavity lines are due to the coupling between the first diffraction orders and the zeroth diffraction orders\cite{Feis2019} (Fabry-Perot modes). The latter modes feature helicity flipping perpendicular reflections between the cavity walls (see Fig.~\ref{fig:single}(d)). When the $(L,f)$ modal condition lines of those Fabry-Perot modes cross a first diffraction order line, a mixed mode forms if the differential phase that the two modes accumulate in a half round-trip is an integer multiple of $2\pi$ \cite{Chang-Hasnain2012,Feis2019}. The mixed modes loose the helicity preserving properties of the grazing first order modes, which causes a sharp degradation of the helicity purity of the field, severely disrupting the enhancement. In the single-array cavity, the two half round-trips of a given mode are not equivalent and, as a consequence, the condition for the creation of a mixed mode is not fulfilled. Ultimately, this is due to the lack of parity symmetry ($\mathbf{r}\rightarrow-\mathbf{r}$) of the single-array cavity. The substitution of one of the disk arrays with the silicon slab breaks the parity symmetry of the double-array cavity. One mirror symmetry plane is also lost. This symmetry breaking is a consequential difference between the two cavity designs. 

The presence of reduced incidence angles due to the silicon slab is another consequential difference. Due to the difference between the refractive indices of the solvent and the silicon film, the propagation of light inside the silicon slab occurs at angles that are much smaller than in the solution. Then, the reflection off the silicon-substrate interface near the bottom of Fig.~\ref{fig:single}(c) is never near-grazing anymore for the first diffraction orders. For example, when the angle of incidence onto the solution-silicon interface tends to 90 degrees, the angle of incidence onto the silicon-substrate interface tends to 25.1 degrees. The small angles of incidence cause the splitting of the TE- and TM- polarizations, degrading the helicity preservation of the reflections and hence of the modes. This splitting is behind the other salient difference in Fig.~\ref{fig:double}: Each of the single enhancement lines in the double-array cavity splits into two lines in the single-array cavity. 

We now investigate this difference in more detail. To avoid duplicity in the discussions, we will from now on focus on TCD, which is typically simpler to measure than ACD. Figures~\ref{fig:gotit}(a,c) show frequency cuts out of the plots in Figs.~\ref{fig:double}(a,b) at $L$=\SI{21.52}{\micro\meter}. The enhancements are now shown on a linear rather than on a logarithmic scale. For the single-array cavity, Fig.~\ref{fig:gotit}(a) shows that all the modal enhancement lines are split, including the lowest frequency mode as seen in the inset, and that their splitting grows as the frequency increases. The same is true for the double-array cavity in Fig.~\ref{fig:gotit}(c) except that the lowest frequency mode is not split. To explain these observations we consider the frequency responses $\alpha_{\text{TE}}^\omega$ and $\alpha_{\text{TM}}^\omega$ of a pair of TE and TM resonances with the same amplitude and separated by a frequency splitting $\Delta \omega$. We assume that they both have a Lorentzian lineshape with linewidth $\gamma_c$. The helicity preservation ($h_c^\omega$) and helicity flip ($h_f^\omega$) response of the combination of resonances is 
\begin{equation}
	\label{eq:hchf}
	h_c^\omega=\frac{\alpha_{\text{TE}}^\omega+\alpha_{\text{TM}}^\omega}{\sqrt{2}}, \ h_f^\omega=\frac{\alpha_{\text{TE}}^\omega-\alpha_{\text{TM}}^\omega}{\sqrt{2}},
\end{equation}
respectively. Figures~\ref{fig:gotit}(b,d) show the normalized helicity content ($|h_c^\omega|^2-|h_f^\omega|^2$) for different normalized separations $\Delta \omega/(2\pi\gamma_c)$ as a function of the normalized angular frequency $\omega/\omega_0$, where $\omega_0$ is the mean of the two resonance frequencies. The helicity content is normalized to the maximum value taken by $|h_c^\omega|^2-|h_f^\omega|^2$ in the case $\Delta \omega=0$. In achiral structures, the helicity content is a good proxy for the CD signal\cite{Graf2019}. For the larger $\Delta \omega/(2\pi\gamma_c)$ values in Fig.~\ref{fig:gotit}(c), the helicity is preserved, flipped, and then preserved again as the frequency changes. This sequence can be understood considering \Eq{eq:hchf} and the change of $\pi$ that the phase of the individual TE (TM) response undergoes when crossing its resonance. As $\Delta \omega/(2\pi\gamma_c)\rightarrow0$, the two resonances progressively degenerate into a single polarization preserving resonance. The ability of the model to reproduce the different mode line structures in Figs.~\ref{fig:gotit}(a,b) strongly suggests the following explanation. The TE/TM degeneracy in the model at $\Delta \omega/(2\pi\gamma_c)=0$ corresponds to the limit of reflection at 90 degrees with respect to the surface normal inside the cavities. In both cavities, such degeneracy is progressively degraded by the increase of the incidence angle of the first order modes as the frequency increases for a given $L$ (Eq.~S10 in the Supp. Mat. of Ref.~\onlinecite{Feis2019}): The incidence angles inside the cavity for the modes in Figs.~\ref{fig:gotit}(a,b) are approximately 85, 80, 75, and 71 degrees (computed with Eq.~(B9) in Ref.~\onlinecite{Feis2019}). The last mode is not visible in Fig.~\ref{fig:gotit}(a). As the frequency decreases, the shapes in Figs.~\ref{fig:gotit}(a,c) can be reproduced by the model using an increasing $\Delta \omega/(2\pi\gamma_c)$. In the case of the single-array cavity, small incidence angles are inherently present due to the silicon slab and prevent $\Delta \omega/(2\pi\gamma_c)$ to become small enough to produce a single helicity preserving peak. The latter actually happens for the first mode of the double-array cavity, as seen in the inset. Additionally, the coupling to the Fabry-Perot modes is another factor that can contribute to the splitting in the double-array cavity, for the reasons discussed above. 
\begin{figure*}[ht!]
    \hspace*{-0.5cm}
    \begin{subfigure}[h!]{.45\linewidth}
    \centering
    \includegraphics[height=6cm]{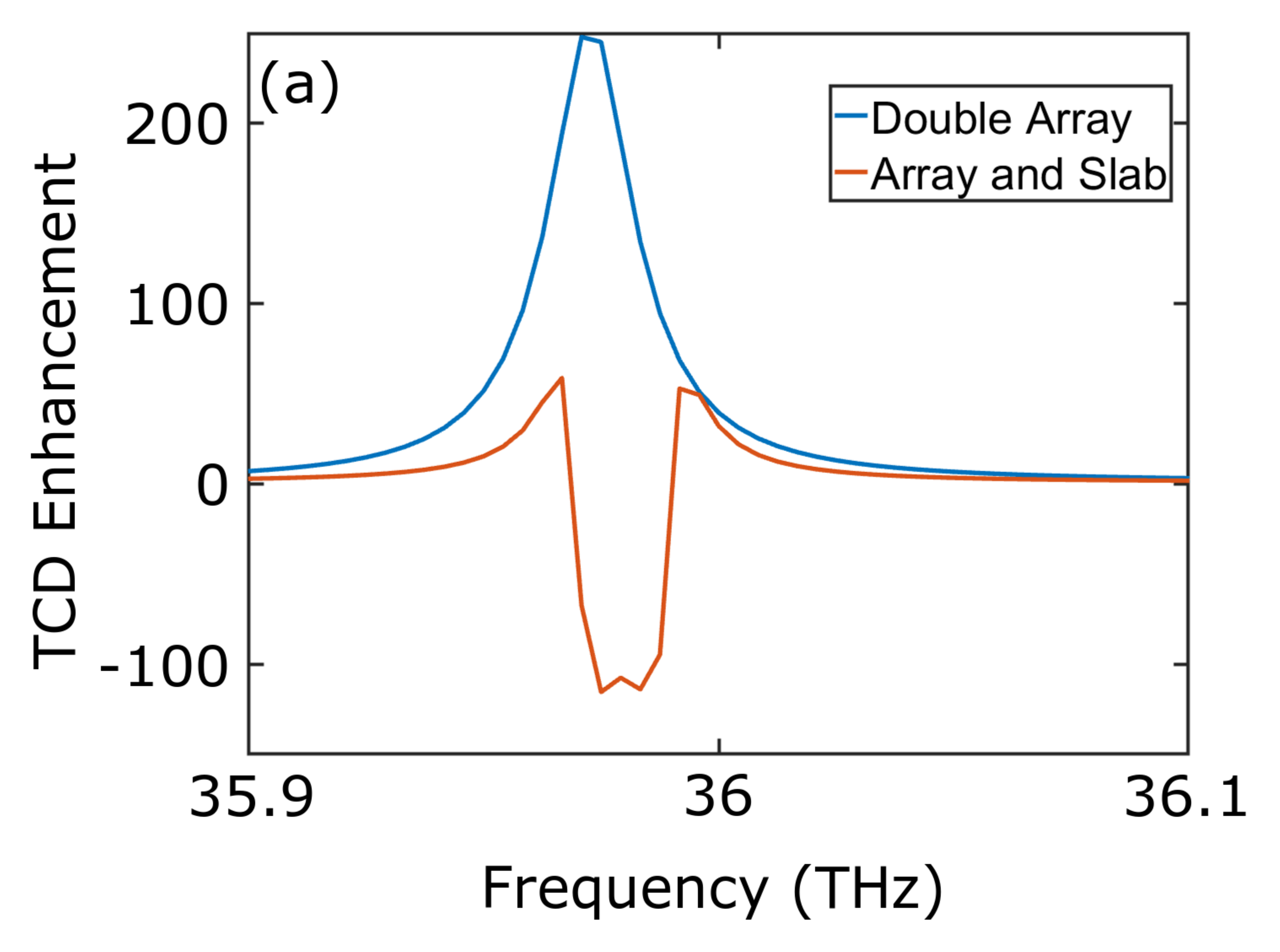}
	\end{subfigure}
	\begin{subfigure}[h!]{.45\linewidth}
    \centering
    \includegraphics[height=6cm]{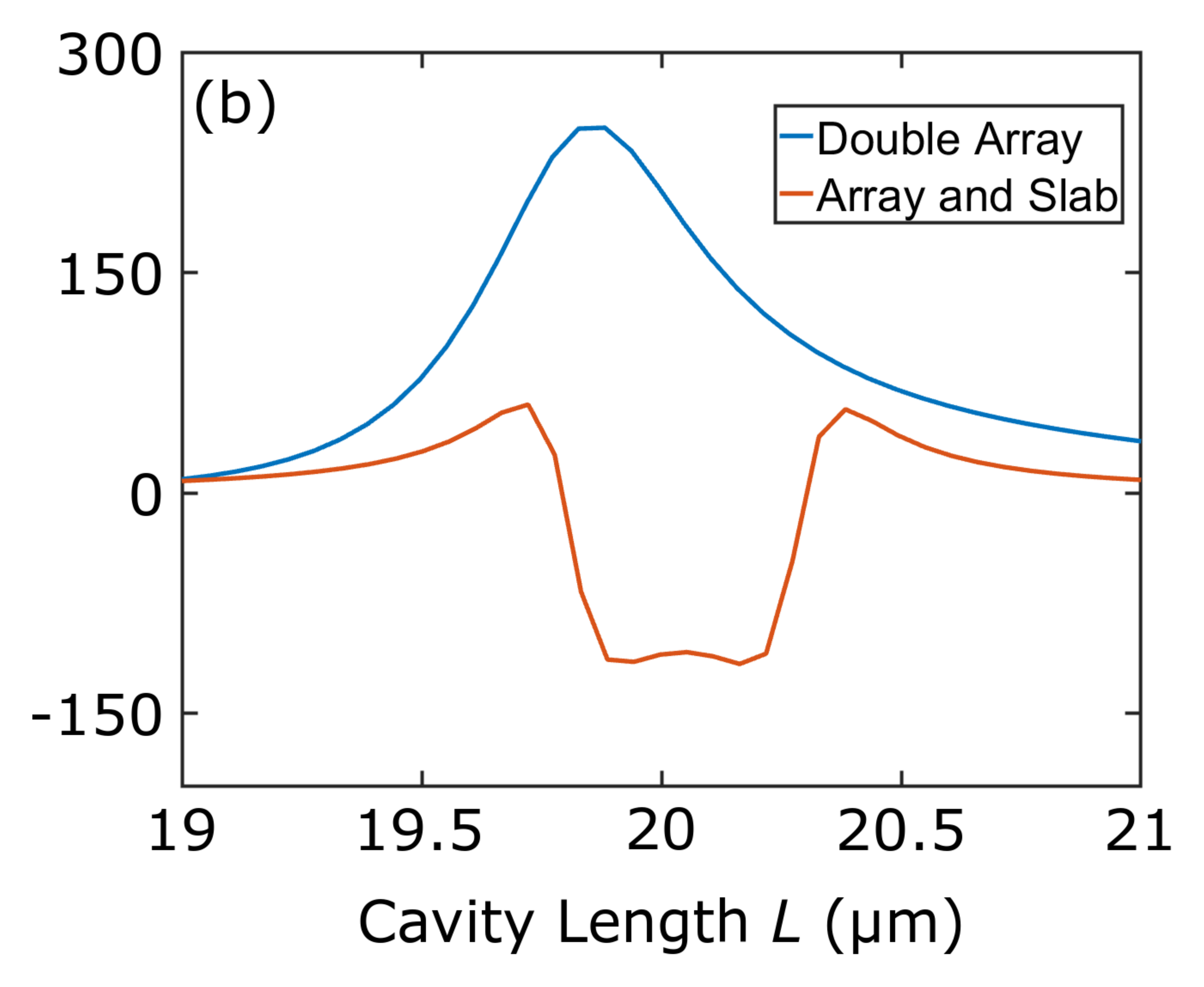}
    \end{subfigure}
	\caption{\label{fig:compar} Plots of the data cuts through the red crosses in Figs.~\ref{fig:double}(a,b) comparing the TCD enhancement (in linear units here) of the double-array cavity design with the simplified single-array and silicon slab design. The cuts marked in the figure are centered at $L$=\SI{19.99}{\micro\meter} and $f$=\SI{35.97}{\tera\hertz}. This figure suggests that the single-array cavity can be operated in two ways for chiral molecule sensing: Using one of the two positive enhancement peaks or using the negative enhancement region in between them, followed by inverting the resulting CD signal.
	}
\end{figure*}

\subsection{Operational considerations}
Once the structure of the enhancement lines has been understood, let us now concentrate on a particular region and consider some practical aspects of the operation of the cavities. Figure~\ref{fig:compar} shows cuts out of the plots in Fig.~\ref{fig:double} in both frequency and cavity length. The cuts are marked with red crosses in Figs.~\ref{fig:double}(a,b), which are centered at $L$=\SI{19.99}{\micro\meter} and $f$=\SI{35.97}{\tera\hertz}. The enhancements are shown in linear units. Figure~\ref{fig:compar} and Figs.~\ref{fig:gotit}(c,d) suggest that the single-array cavity could be operated in two different ways for sensing chiral molecules: Using one of the two positive enhancement peaks, or using the negative enhancement region in between them and inverting the resulting CD signal. The linewidths of lasers and the resolution of spectrometers\cite{Jasco2020} in this frequency region suggest that the operation using the area of negative enhancement is straightforward, and that selecting the positive peaks is more challenging because their width is comparable to the resolution of standard spectrometers of $\approx$\SI{0.015}{\tera\hertz}. Such resolution is much finer than the width of the enhancement line of the double-array cavity shown in Fig.~\ref{fig:compar}(a). Regarding the cavity length $L$, the precision needed to select each peak is experimentally feasible using linear actuators with high precision. While the narrower peaks in Fig.~\ref{fig:compar}(b) have a linewidth of about \SI{0.15}{\micro\meter}, there are commercially available linear actuators with a minimal incremental motion of \SI{0.05}{\micro\meter}. Nevertheless, the simulations assume that the slab and the array are perfectly parallel to each other. As discussed above, the degree of parallelism between the two sides of a cavity can be controlled by using a set of three linear actuators in a triangular formation at the edges of the cavity to adjust one of the sides (see the inset in Fig.~\ref{fig:CDspec}). While the numerical simulation of a residual inclination between the two sides is challenging, its effects are likely to be akin to those due to the cavity length varying across the transverse dimensions. The inclination should hence be carefully controlled to avoid averaging across the positive and negative enhancements seen in Fig.~\ref{fig:compar}(b) for the single-array cavity. The uniform sign of enhancement in the double-array cavity makes it more robust against this. Finally, we note that both cavity designs can acquire some degree of chirality depending on the kind of inclination. The end effect of this chirality needs to be experimentally investigated. Overall, the considered operational requirements are more stringent for the single-array cavity. This partially counteracts the benefits of such simplified design, which does not require the lateral alignment of two disk arrays with deep sub-micrometer precision.

Finally, we foresee the fabrication of silicon-disk arrays with a lateral extent close to \SI{20}{\milli\meter}. For a length $L$=\SI{20}{\micro\meter}, the cavity would hold \SI{0.8}{\milli\liter} of analyte.

We will now address the realistic modeling of chiral molecular resonances which is needed to accurately predict the performance of our CD enhancing cavities.
\section{Chiral molecule resonance model with experimental input\label{sec:two}}
Constitutive relations of the type written in \Eq{eq:cons} are routinely assumed for the numerical evaluation of CD enhancement systems \cite{Nesterov2016,Mohammadi2018,Garcia-Guirado2018,Garcia-Guirado2020,Graf2019,Feis2019,Droulias2020}. To the best of our knowledge, a frequency-dispersive model in accordance with energy conservation [Eq.~(3.219) in Ref.~\onlinecite{Lindell1992}] and fully determined by available CD measurements has not yet been reported. Such a model is needed to elucidate possible effects due to the interplay between the resonances of the molecules and the resonances of the sensing system. Even when the difference in the lifetimes of the two kinds of resonances is large, suggesting a correspondingly small energy exchange rate between them, the use of realistic parameters is important because effects that would otherwise be negligible can play a significant role in systems that achieve an unusually large light-molecule interaction time, like our cavity designs. In particular, the settings that we used in Ref.~\onlinecite{Feis2019} $\varepsiloneff=(2.14+\ii10^{-4})\varepsilon_0$, $\mueff=(1+\ii10^{-4})\mu_0$, and $\kappaeff=\ii 10^{-4}$, meet energy conservation but also implicitly assume that the molecules have an unrealistic degree of duality symmetry, that is, that the handedness of the incident light changes much less during the light-molecule interaction than for real chiral molecules. A foreseeable effect of the helicity change is the progressive degradation of the polarization purity of the light in the molecular solution. Intuitively, the CD signal decreases with decreasing polarization purity and vanishes in the limit when the light is a perfect mix of both helicities. More surprisingly, we will show in the next section that the lack of duality symmetry of the molecules can significantly disturb the cavity modes. None of these effects will be noticeable when the interaction time between the light and the molecules is much smaller that the inverse of the rate of change of helicity in the solution, like in typical CD spectrometers or non-cavity based resonant CD enhancement systems. We now present the mentioned model. 

We consider a homogeneous and isotropic solution of identical chiral molecules. Our objective is to describe the frequency-dependent constitutive relations of such medium in spectral proximity to a resonance of the molecule. To such end, we will consider the constitutive relations in \Eq{eq:cons} and find expressions for the scalars $\varepsiloneff$, $\mueff$, and $\kappaeff$ by means of a microscopic model and its subsequent homogenization. The microscopic light-molecule interaction description that we consider is the molecular linear polarizability matrix
\begin{equation}
	\label{eq:alphamat}
	\inddipoles=\begin{bmatrix}\aee&\aem\\\ame&\amm\end{bmatrix}\fields=\alphamat\fields,
\end{equation}
where the electromagnetic field oscillating at frequency $\omega$ incident on a molecule at position $\rr$ induces electric and magnetic dipolar moments $\dd$ and $\mm$, respectively, according to a 6$\times$6 complex matrix $\alphamat$. This matrix fully determines the linear light-molecule interaction to dipolar order, including both scattering and absorption of the incident light. In our case, the random orientation of the molecules in the solution justifies the assumption of isotropy at the microscopic level. This means that each of the $\alpha_{xy}^\omega$ in \Eq{eq:alphamat} can be assumed to be the 3$\times$3 identity matrix multiplied by a complex scalar, which we denote by the same $\alpha_{xy}^\omega$ symbols in a slight abuse of the notation that we adopt from now on.

We now assume the quasi-static model for the dipolar resonance of a chiral object from Eq.~(20) in Ref.~\onlinecite{Sersic2011}, which, after imposing isotropy reads
\begin{equation}
	\alphamat=\begin{bmatrix}\aee&\aem\\\ame&\amm\end{bmatrix}=\frac{V}{1-\baromega^2+\ii\bargamma\baromega}\begin{bmatrix}\etaee&\ii\etaem\\-\ii\etame&\etamm\end{bmatrix},
\end{equation}
where the $\eta_{xy}$ are real, semi-positive, unitless, and frequency-independent, $\baromega=\frac{\omega}{\omega_0}$, $\bargamma=\frac{\gamma}{\omega_0}$, and $\omega_0$ and $\gamma$ are the resonance frequency and linewidth of the chiral molecule resonance, respectively. The factor $V$ has units of volume in the natural units of Tab.~I in Ref.~\onlinecite{Sersic2011}. Additionally, we use the condition in Eq.~(6.126) of Ref.~\onlinecite{Lindell1994}
\begin{equation}
	\label{eq:cable}
\etaee\etamm=\etaem^2,
\end{equation}
which is met when the electric and magnetic dipolar moments originate from the same current distribution. Then, after defining $\beta=\frac{\etamm}{\etaee}$, we can write
\begin{equation}
	\label{eq:nonsi}
	\alphamat=\frac{V\etaee}{\den}\begin{bmatrix}1&\ii s\sqrt{\beta}\\-\ii s\sqrt{\beta}&\beta\end{bmatrix},
\end{equation}
where $s=\{-1,+1\}$ is a sign factor that determines the handedness of the resonance. Opposite molecular enantiomers will feature opposite signs. Note that, our assumptions imply that $\beta\ge 0$. When $\beta=0$, the resonance is purely electric and there is no chiral response. A change to the helicity basis for both the electromagnetic fields and the induced dipoles \cite{FerCor2013} shows that when $\beta=1$ the resonance is excited only by electromagnetic fields of a fixed helicity determined by $s$, and that, upon interaction, the helicity of the scattered field is the same as that of the incident field. When $\beta=1$ the resonance is hence maximally electromagnetically chiral \cite{FerCor2016} and dual symmetric \cite{FerCor2013}. Values of $\beta>1$ correspond to chiral resonances whose magnetic component is larger than the electric component, and $\beta\rightarrow\infty$ to a purely magnetic resonance. 

Equation~(\ref{eq:nonsi}) can be changed to SI units using Tab.~I in Ref.~\onlinecite{Sersic2011}: 
\begin{equation}
	\label{eq:alphamatmodel}
	\alphamat=\frac{4\pi V\etaee}{\den}\begin{bmatrix}\varepsilon&\ii s\frac{\sqrt{\beta}}{c}\\-\ii s\frac{\sqrt{\beta}}{c}&\beta\end{bmatrix},
\end{equation}
where $\varepsilon=\varepsilonr\varepsilon_0$ is the electric permittivity of the solvent, which we assume to be non-dispersive in the vicinity of the resonance. The solvent is also assumed to be achiral and non-magnetic ($\kappa=0,\ \mu=\mu_0$). The speed of light in the solvent is $c=1/\sqrt{\varepsilon\mu}$. 

At this point, the common homogenization equations for dilute mixtures of small chiral inclusions embedded in a dielectric media [see e.g. Eq.~(6.127) in Ref.~\onlinecite{Lindell1994}] can be used\footnote{We ignore the factor $D$ in Eq.~(6.127) of Ref.~\onlinecite{Lindell1994} because in our case it is much smaller than 1.} to obtain the following expressions for the quantities in \Eq{eq:cons}
\begin{equation}
	\label{eq:finalcons}
	\begin{split}
		\varepsiloneff&=\varepsilon\left(1+\frac{4\pi}{3}\frac{\hatV}{\den}\right)\\
		\mueff&=\mu\left(1+\frac{4\pi}{3}\frac{\hatV\beta}{\den}\right),\\
		\kappaeff&=\frac{4\pi}{3}\frac{\hatV s\sqrt{\beta}\sqrt{\varepsilonr}}{\den}.
	\end{split}
\end{equation}
The unitless parameter $\hatV=\rho V\etaee$ collects all the factors that determine the magnitude of the influence of the molecules: $V$ and $\etaee$ from the microscopic resonance model, and the macroscopic concentration of molecules $\rho$. The parameter $\hatV$ can hence be varied for the study of a given CD enhancement system. We will now show how all the other parameters of the microscopic model that appear in \Eq{eq:finalcons}, namely $s$, $\beta$, and $\bargamma$, can be obtained from existing experimental measurements.

In typical CD spectrometers, a fixed volume of the molecular solution is subsequently illuminated by propagating light beams of opposite polarization handedness ($\pm$) covering a certain frequency bandwidth. The absorption $A_{\pm}^\omega$ in each case is measured as a function of the frequency. Their average and difference are typically reported as a function of the frequency, as well as the CD: 
\begin{equation}
	\label{eq:sigmadelta}
	\Sigma A^\omega=\frac{A_{+}^\omega+A_{-}^\omega}{2},\ \Delta A^\omega= A_{+}^\omega-A_{-}^\omega,\ \mathrm{CD}^{\omega}=\frac{\Delta A^{\omega}}{\Sigma A^{\omega}}.
\end{equation}
With this definition, the absolute value of the $\mathrm{CD}^{\omega_0}$ is equal to the Kuhn's dissymmetry factor \cite{Kuhn1930}. There is a large amount of chiral spectroscopic measurements available in the literature covering many different molecules and proteins \cite{Abbate2012,Mazzeo2015,Moreno2009,Ortega2015,Quesada2018}. Let us assume that we are interested in a particular resonance of a particular chiral molecule. The parameters $\omega_0$ and $\bargamma=\gamma/\omega_0$ in the model can immediately be determined from the central frequency and linewidth of the available CD measurements of the molecular resonance. To show how $s$ and $\beta$ are determined, we take again a microscopic point of view and consider a molecule at point $\rr$. Using Eq.~(5) in Ref.~\onlinecite{Graf2019} and \Eq{eq:alphamatmodel}, the resonant absorption of the molecule at $\omega=\omega_0$ can be written as
\begin{equation}
	\label{eq:apm}
	A_{\pm}^{\omega_0}(\rr)=|\mathbf{G}^{\omega_0}_\pm(\rr)|^2\frac{2\pi\omega_0\varepsilon V\etaee}{\bargamma}\left(\frac{1+\beta}{2}\pm s\sqrt{\beta}\right).
\end{equation}
where $\sqrt{2}\mathbf{G}^{\omega_0}_\pm(\rr)=\mathbf{E}^{\omega_0}(\rr)\pm \ii Z\mathbf{H}^{\omega_0}(\rr)$ are a version of the Riemann-Silberstein vectors\cite{Birula1996,Birula2013}. The experimentally measured absorbances $A_{\pm}^{\omega_0}$ are obtained after integrating \Eq{eq:apm} on the appropriate volume, and accounting for the molecular concentration. Then, using Eqs.~(\ref{eq:sigmadelta}) and (\ref{eq:apm}) it is straightforward to show that
\begin{equation}
	\label{eq:CDbeta}
	\mathrm{CD}^{\omega_0}=\frac{\Delta A^{\omega^0}}{\Sigma A^{\omega^0}}=s\frac{4\sqrt{\beta}}{1+\beta},
\end{equation}
which determines $s$ and $\beta$
\begin{equation}
	\label{eq:betacd}
	s=\text{sign}\left\{\mathrm{CD}^{\omega_0}\right\},\ \beta=\left[\frac{2}{|\mathrm{CD}^{\omega_0}|}-\sqrt{\frac{4}{|\mathrm{CD}^{\omega_0}|^2}-1}\right]^2.
\end{equation}

In this way, the resonance specific parameters $\omega_0$, $\bargamma$, $s$, and $\beta$ are all fixed by experimental data, and only $\hatV$ is left free. For our later purposes, it is convenient to relate $\hatV$ to the modulus of the constitutive chiral parameter at the resonance frequency $|\kappa_{\mathrm{eff}}^{\omega_0}|$. From the last line of \Eq{eq:finalcons} we obtain that 
\begin{equation}
	\label{eq:Vkappa}
	\hatV=\frac{3}{4\pi}\frac{|\kappa_{\mathrm{eff}}^{\omega_0}|\bargamma}{\sqrt{\beta}\sqrt{\varepsilonr\mur}}.
\end{equation}

\begin{figure*}[ht!]
    \centering
    \includegraphics[width=\textwidth]{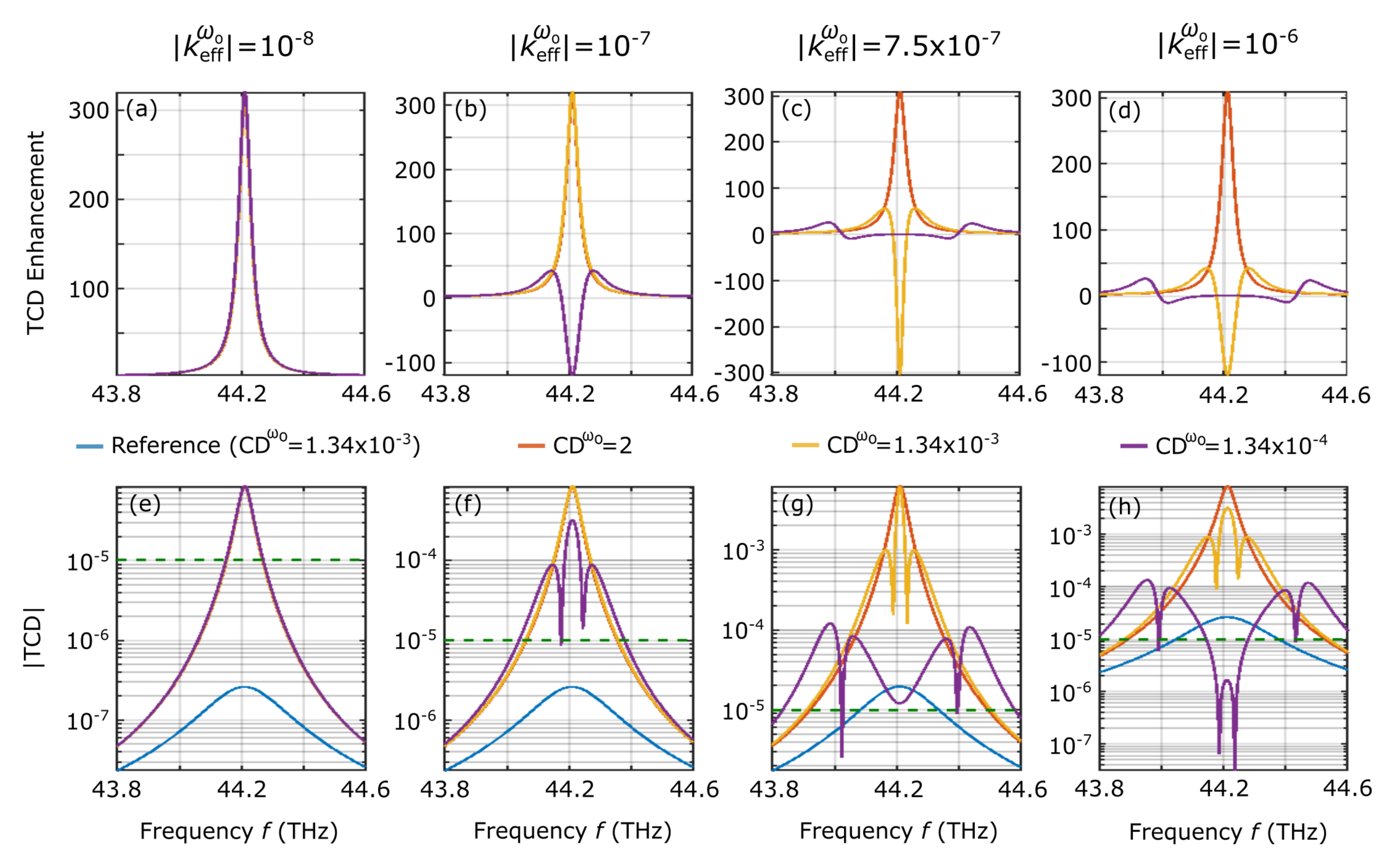}
	\caption{\label{fig:resonance} Results for the double-array cavity. Panels (a)-(d) show the TCD enhancement for different values of $|\kappa_{\mathrm{eff}}^{\omega_0}|$ (see titles). In each of the panels, the TCD enhancement for different values of $\mathrm{CD}^{\omega_0}$ are shown (see legend). Panels (e)-(h) show the absolute value of the TCD for the reference case without the arrays (blue lines) and for the cavity (red, yellow, and purple lines). The green-dashed line marks a detection threshold (see text).}
\end{figure*}
\begin{figure*}[ht!]
    \centering
    \includegraphics[width=\textwidth]{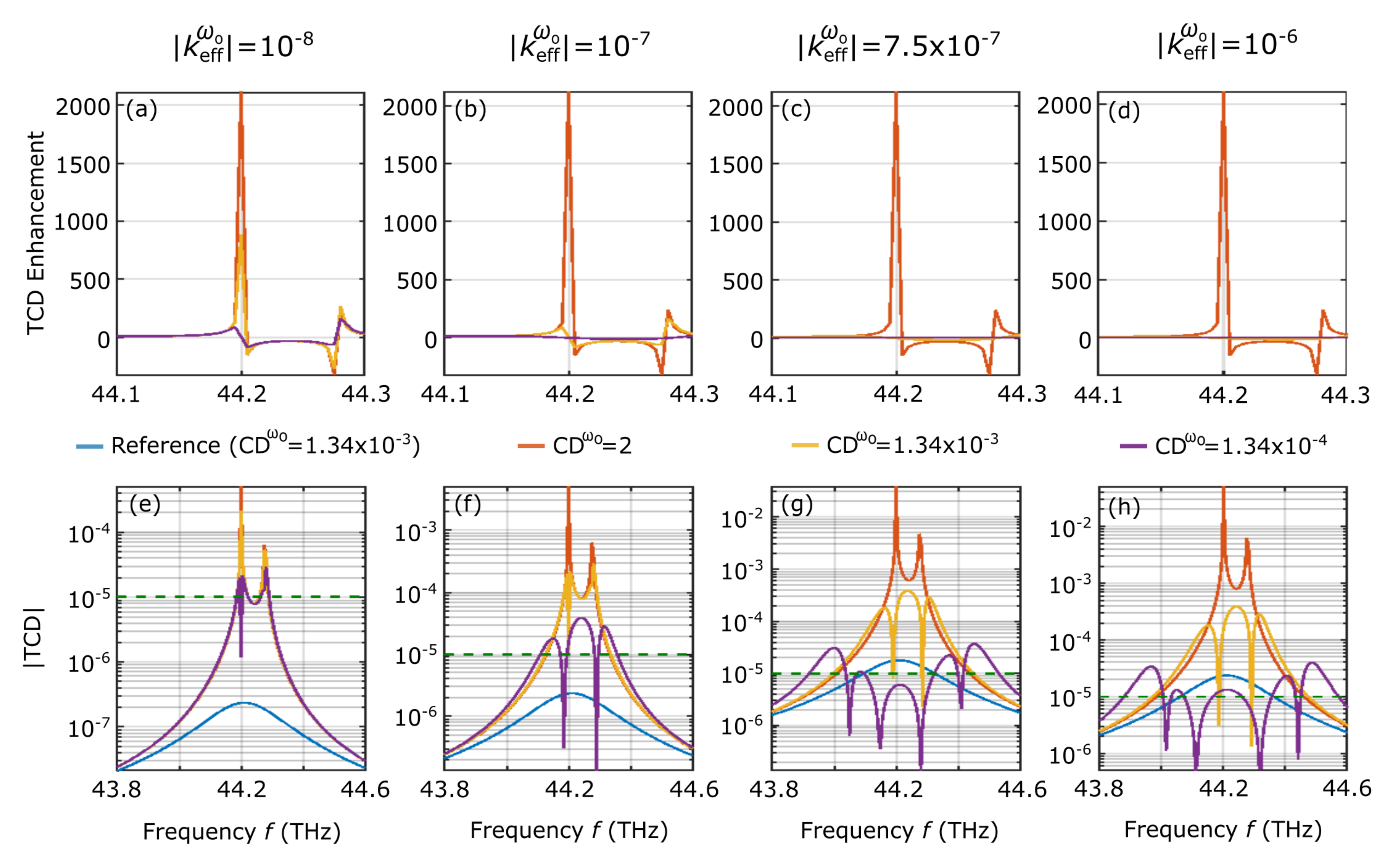}
	\caption{\label{fig:resonanceS} Results for the single-array cavity. Panels (a)-(d) show the TCD enhancement for different values of $|\kappa_{\mathrm{eff}}^{\omega_0}|$ (see titles). For improved visibility, the frequency range on the horizontal axes of the top panels is a zoomed-in portion of the frequency range in the bottom panels. In each of the panels, the TCD enhancement for different values of $\mathrm{CD}^{\omega_0}$ are shown (see legend). Panels (e)-(h) show the absolute value of the TCD for the reference case without the arrays (blue lines) and for the cavity (red, yellow, and purple lines). The green-dashed line marks a detection threshold (see text).}
\end{figure*}

\section{Sensing of chiral molecules in a cavity\label{sec:three}} 
We now use the model to predict the TCD enhancement and the TCD signals of the different cavities for a particular resonance of binol. We use the VCD measurements from a commercial spectrometer (Fig.~4 in Ref.~\onlinecite{Jasco2011}) to extract the following parameters: $\omega_0/(2\pi)=f_0=$\SI{44.21}{\tera\hertz}, $\gamma=$\SI{0.258}{\tera\hertz}, $\mathrm{CD}^{\omega_0}=1.34\times 10^{-3}$, $s=+1$, and obtain $\beta=1.12\times 10^{-7}$. We also adapt the following parameters of the cavity to the target frequency [see Figs.~\ref{fig:single}(b,c)]: $h=$\SI{1.082}{\micro\meter}, $r=$\SI{1.574}{\micro\meter}, $a=$\SI{4.721}{\micro\meter}, and a relative permittivity equal to 1.8912 for the substrate and the solvent. We use the lowest frequency mode in both cavities and a cavity length equal to $L=$\SI{11.52}{\micro\meter}(\SI{12.80}{\micro\meter}) for the single(double) array cavity. We consider four different values of $|\kappa_{\mathrm{eff}}^{\omega_0}|\in [10^{-8},10^{-7},7.5\times 10^{-7},10^{-6}]$, which imply four different concentrations $\rho$ proportional to each value of $|\kappa_{\mathrm{eff}}^{\omega_0}|$ according to Eqs.~(\ref{eq:Vkappa},\ref{eq:finalcons}). Besides binol (yellow lines in Figs.~\ref{fig:resonance},\ref{fig:resonanceS} with $\mathrm{CD}^{\omega_0}=1.34\times 10^{-3}$), we also analyze the results for $\mathrm{CD}^{\omega_0}=1.34\times 10^{-4}$ (purple lines) and $\mathrm{CD}^{\omega_0}=2$ (red lines), which allow us to study the effects of different degrees of duality breaking (helicity change) of the molecules. It can be deduced from \Eq{eq:betacd} that the perfect duality condition $\beta=1$ is achieved when $\mathrm{CD}^{\omega_0}=\pm 2$, and that the departure from duality increases as $|\mathrm{CD}^{\omega_0}|$ decreases. 

Figures~\ref{fig:resonance} and \ref{fig:resonanceS} show the TCD enhancement and the TCD signal for the double-array and single-array cavity, respectively. Let us first analyze the results for the double-array cavity. We start with the perfectly dual $\mathrm{CD}^{\omega_0}=2$ case. The plots in Figs.~\ref{fig:resonance}(a)-(d) show that the enhancement is independent of the concentration. In particular, the maximum enhancement of $\approx 310$ is always achieved at the molecular resonance frequency. Additionally, the plots of the absolute value of the TCD signal in Figs.~\ref{fig:resonance}(e)-(h) are identical to each other except for the expected scaling with $|\kappa_{\mathrm{eff}}^{\omega_0}|$. This shows that the effect of the cavity is independent of the molecular concentration if there is no helicity change in the solution. For $\mathrm{CD}^{\omega_0}=2$ the cavity is always working in an undisturbed fashion. In contrast, the TCD enhancement and the spectral signature of the TCD signal depend on $|\kappa_{\mathrm{eff}}^{\omega_0}|$ for the non-dual cases ($\mathrm{CD}^{\omega_0}\neq 2$). The previously discussed reduction of the CD signal due to the light changing helicity upon interaction with the molecules can at most bring the enhancement to zero and does not explain the sign changes that we observe in Figs.~\ref{fig:resonance}(b-d). These can nonetheless be explained by noticing that the shapes of the enhancement lines in Figs.~\ref{fig:resonance}(b-d) are essentially the same as those in Figs.~\ref{fig:gotit}(c,d), indicating that the helicity change due to the molecules causes a split of the TE and TM resonances of the cavity. We note that the undisturbed cavity resonance for the double-array cavity is always helicity preserving,  as can be deduced from the results with lowest $|\kappa_{\mathrm{eff}}^{\omega_0}|$ in Fig.~\ref{fig:resonance}(a). The plots in Fig.~\ref{fig:resonance} suggest that the splitting is a function of $\rho/\mathrm{CD}^{\omega_0}$. For example, the shape changes that can be seen for $\mathrm{CD}^{\omega_0}=1.34\times 10^{-3}$ when going from $|\kappa_{\mathrm{eff}}^{\omega_0}|=10^{-7}$ to $|\kappa_{\mathrm{eff}}^{\omega_0}|=10^{-6}$, are identical to the changes that can be seen for $\mathrm{CD}^{\omega_0}=1.34\times 10^{-4}$ when going from $|\kappa_{\mathrm{eff}}^{\omega_0}|=10^{-8}$ to $|\kappa_{\mathrm{eff}}^{\omega_0}|=10^{-7}$. 

A similar analysis holds for the corresponding results for the single-array cavity in Fig.~\ref{fig:resonanceS}, except that the undisturbed single cavity resonance is already split, as shown by the $\mathrm{CD}^{\omega_0}=2$ lines. Then, there is an additional split which increases with $\rho/\mathrm{CD}^{\omega_0}$. Another difference is that the maximum enhancement ($\approx 2100$ for the $\mathrm{CD}^{\omega_0}=2$ case), is never achieved in the case of binol, whose largest positive enhancement is $\approx 850$ for the lowest value of $|\kappa_{\mathrm{eff}}^{\omega_0}|$. On the other hand, the region of negative enhancement, featuring an enhancement of $\approx -35$ at its center, is very similar for binol and for the perfectly dual case when $|\kappa_{\mathrm{eff}}^{\omega_0}|\le 10^{-7}$. These observations are consistent with $\rho/\mathrm{CD}^{\omega_0}$ having a larger (smaller) disturbing effect for thiner (wider) enhancement linewidths, i.e. larger (smaller) light-matter interaction times.

Let us now further discuss the absolute value of the TCD measurements shown in Figs.~\ref{fig:resonance},\ref{fig:resonanceS}(e-h) for the case of binol. Whether the actual TCD value is above or below the measurement detection threshold determines the feasibility of the sensing experiment. For discussion purposes, we assume a sensitivity threshold of 10$^{-5}$, which is close to the 8$\times$10$^{-6}$ noise level of some commercially available CD spectrometers \cite{Nafie2011,Jasco2020}. The threshold is drawn as a green-dashed line in the bottom rows of Figs.~\ref{fig:resonance},\ref{fig:resonanceS}. For $|\kappa_{\mathrm{eff}}^{\omega_0}|\le 10^{-7}$, the reference signal without the cavities (blue line) is well below the detection threshold. In these cases, both cavities are capable of bringing the TCD signal above the threshold.

We note that the chiral resonance model can be easily extended to solutions containing both enantiomers of the molecule. It suffices to modify the values of the concentration $\rho$ and of the CD
\begin{equation}
	\rho \rightarrow \rhoL+\rhoD,\ \mathrm{CD}^{\omega_0}\rightarrow \frac{\rhoL-\rhoD}{\rhoL+\rhoD}|\mathrm{CD}^{\omega_0}|,
\end{equation}
where $\rhoL(\rhoD)$ is the concentration of the L(D) mirror version of the chiral molecule and $\frac{\rhoL-\rhoD}{\rhoL+\rhoD}$ is commonly known as the enantiomeric excess (ee). 

Finally, we note that the CD enhancing cavities can also work as optical rotation enhancing systems after appropriate changes in the illumination and measurements. Preliminary simulation results show that the OR enhancements are similar to the discussed CD enhancements, which are smaller than the four orders of magnitude OR enhancements predicted for the multi-pass CRDP technique \cite{Bougas2014}. Also, at their current sizes, CRDP setups offer an advantageous frequency-independent enhancement because their free-space cavity lengths are longer than the coherence length of light. This prevents the frequency-dependent constructive and destructive interference that leads to the typical mode build-up in shorter cavities.

\section{Concluding remarks\label{sec:four}}
We have numerically demonstrated that different optical cavities can resonantly enhance the vibrational circular dichroism (VCD) signal of solutions of chiral molecules by one to three orders of magnitude, for a given molecule concentration and given thickness of the cell containing the molecules. In particular, we have compared a newly introduced simplified cavity based on a single silicon disk array and a silicon film with a previously introduced design featuring two arrays of silicon disks. In this article, we have used a realistic model for the electromagnetic response of the solution of chiral molecules at the frequency region containing a molecular resonance. This refined model has allowed us to uncover a rather surprising effect: The structure of the cavity modes can be affected by the lack of duality symmetry of the molecules, that is, by the partial change of the handedness of light upon light-molecule interaction. For sufficiently large ratios of the molecular concentration divided by the Kuhn's dissymmetry factor of the molecular resonance, the duality breaking of the molecules causes a split of the TE and TM components of the cavity modes. This split is in addition to other TE/TM splits that are inherent to the design of the cavities, and disturb the ideal degeneracy featured by perfectly helicity-preserving modes.

The experimental implementation of our CD enhancement cavities should extend the capabilities of traditional chiral spectrometers to unprecedentedly small volumes of analyte. In this respect, substituting the sample cell inside CD spectrometers by our cavities is a straightforward implementation path which is currently possible, and that can be used as a stepping stone towards applications on future lab-on-a-chip devices.

\begin{acknowledgments}
	This research has been funded by the Hector Fellow Academy, by the Deutsche Forschungsgemeinschaft (DFG, German Research Foundation) under Germany's Excellence Strategy via the Excellence Cluster 3D Matter Made to Order (EXC-2082/1 -- 390761711) and via the SFB 1173 (Project-ID 258734477), by the Carl Zeiss Foundation, by the Helmholtz Association via the Helmholtz program ``Materials Systems Engineering" (MSE), and by the KIT through the ``Virtual Materials Design'' (VIRTMAT) project. X.G.-S. is pursuing his Ph.D. within the Karlsruhe School of Optics and Photonics (KSOP) and acknowledges financial support. Finally, we are grateful to the company JCMwave for their free provision of the FEM Maxwell solver JCMsuite. 
\end{acknowledgments}
\section*{Data availability}
The data that supports the findings of this study are available within the article.

\newpage
\end{document}